\documentclass[11pt,a4paper]{article}
\pdfoutput=1
\usepackage{jheppub}
\usepackage{rotating}
\usepackage{array}
\usepackage{amsmath}
\usepackage[normalem]{ulem}
\usepackage{slashed}
\usepackage{booktabs}
\usepackage[pdftex,table]{xcolor}
\usepackage{units}
\usepackage{booktabs}

\newcommand{\be}{\begin{equation}}
\newcommand{\ee}{\end{equation}}
\newcommand{\bea}{\begin{eqnarray}}
\newcommand{\eea}{\end{eqnarray}}
\newcommand{\ds}{{\sf DarkSUSY}}

\usepackage{multirow}
\preprint{CERN-TH-2022-147}

\title{No room to hide: implications of cosmic-ray upscattering for GeV-scale dark matter}

\author[1]{James Alvey}
\author[2,3]{$\!\!$, Torsten Bringmann}
\author[4,5]{$\!\!$, Helena Kolesova}

\affiliation[1]{GRAPPA Institute, Institute for Theoretical Physics Amsterdam,
University of Amsterdam, Science Park 904, 1098 XH Amsterdam, The Netherlands}
\affiliation[2]{Department of Physics, University of Oslo, Box 1048, N-0316 Oslo, Norway}
\affiliation[3]{Theoretical Physics Department, CERN, 1211 Geneva 23, Switzerland}
\affiliation[4]{Department of Mathematics and Physics, University of Stavanger, 4036 Stavanger, Norway}
\affiliation[5]{AEC, Institute for Theoretical Physics, University of Bern, Sidlerstrasse 5, CH-3012 Bern, Switzerland}

\emailAdd{j.b.g.alvey@uva.nl}
\emailAdd{torsten.bringmann@fys.uio.no}
\emailAdd{helena.kolesova@uis.no}

\abstract{%
The irreducible upscattering of cold dark matter by cosmic rays opens up the intriguing possibility of detecting even light dark matter in conventional direct detection experiments or underground  neutrino detectors. The mechanism also significantly enhances sensitivity to models with very large nuclear scattering rates, where the atmosphere and rock overburden efficiently stop standard non-relativistic dark matter particles before they could reach the detector. In this article, we demonstrate that cosmic-ray upscattering essentially closes the window for strongly interacting dark matter in the (sub-)GeV mass range. Arriving at this conclusion crucially requires a detailed treatment of both nuclear form factors and inelastic dark matter-nucleus scattering, as well as including the full momentum-transfer dependence of scattering amplitudes. We illustrate the latter point by considering three generic situations where such a momentum-dependence is particularly relevant, namely for interactions dominated by the exchange of light vector or scalar mediators, respectively, and for dark matter particles of finite size. As a final concrete example, we apply our analysis to a putative hexaquark state, which has been suggested as a viable baryonic dark matter candidate. Once again, we find that the updated constraints derived in this work close a significant part of otherwise unconstrained parameter space.
}

\begin{document}
\maketitle
\flushbottom

\section{Introduction}
\label{sec:introduction}

\noindent The strategies to search for a dark matter (DM) component in the Universe are nowadays extremely varied, targeting 
many possible gravitational and non-gravitational properties such as the DM mass or standard model (SM) 
couplings~\cite{Bertone:2004pz}. In astrophysical, cosmological, and laboratory settings, this broadband approach has yet to 
conclusively reveal any non-gravitational signatures. However, via both indirect and direct searches, the very wide DM 
model space has been significantly restricted. The focus of this article concerns the reach of the generic 
class of experiments aiming to directly detect DM through a possible DM-nucleon coupling~\cite{Goodman:1984dc}, 
known as direct detection facilities. Currently, world-leading examples of this setup include 
e.g.~LUX-ZEPLIN (LZ)~\cite{LZ:2022ufs}, PandaX-4T~\cite{PandaX-4T:2021bab}, and Xenon-1T~\cite{XENON:2018voc},
which set the strongest limits in the DM mass 
$m_\chi$ vs.~spin-independent nuclear coupling $\sigma_{\mathrm{SI}}$ parameter space.

The sensitivity of a given direct detection experiment is controlled by a number of factors. Firstly, the event rate $\Gamma_N$
scales with the number of DM particles that have a sufficiently large kinetic energy. Specifically, the DM energy 
must be large enough 
to induce a nuclear recoil that can trigger a signal above the detector threshold. Secondly, the rate 
also scales linearly with the DM-nucleon cross section $\mathrm{d}\sigma_{\chi N} / \mathrm{d} T_N$, 
at least in the above examples, where $T_N$ is the nuclear recoil energy. Thirdly, as in any count-based experiment, 
this signal rate should be compared to some 
background event rate to derive a statistically significant detection threshold. Notably, in direct detection facilities, the background 
rates are typically extremely low as necessitated by the small expected signal rates, although there are some important 
exceptions, such as a dedicated CRESST surface run~\cite{CRESST:2017ues}. 

The standard target for these experiments is the DM in the Galactic halo, which has characteristic velocities of the order
$v_\chi \sim 10^{-3}c$ and in any case cannot exceed the Galactic escape velocity 
$v_\mathrm{esc} \sim 540 \, \mathrm{km/s}$~\cite{Evans:2000gr,Evans:2005tn}. For a given DM mass 
$m_\chi$, there is hence unavoidably a maximum DM kinetic energy available to excite nuclear recoil signals of the order 
$T_N \sim m_\chi^2 v_\mathrm{esc}^2 / m_N$. For some DM mass $m_\chi^\mathrm{min}$
this recoil energy must fall below the detectable threshold, and the experimental sensitivity drops to zero. For experiments such as 
Xenon, PandaX and LZ, it is well-known that this cut-off
lies around the GeV-scale, corresponding to a detectable threshold in the keV range. As such, even though
these detectors have impressive reach -- currently down to the level of spin-independent cross sections of
$\sigma_\mathrm{SI} \sim 10^{-47}\,\mathrm{cm}^2$~\cite{XENON:2018voc,PandaX-4T:2021bab,LZ:2022ufs}, and even 
approaching the neutrino floor~\cite{Strigari:2009bq,OHare:2021utq} with ongoing searches -- there is ample motivation 
(and hence, in fact, both experimental and theoretical activity)
for methods to probe the sub-GeV mass range~\cite{Knapen:2017xzo,Essig:2022dfa}. 
This describes the first ``window" in which DM can hide -- it could just be that DM has 
a small mass out of the reach of direct detection experiments. 
There is yet another window at {\it large} values of the cross section 
$\sigma_{\mathrm{SI}}$, however, which will be a key focus of this article. This arises due to the fact that if DM interacts \textit{too} 
strongly, then it can actually be the case that DM particles are unable to reach the detectors due to the attenuation of the flux 
in the atmosphere or the rock overburden~\cite{Starkman:1990nj,Zaharijas:2004jv, Mack:2007xj}. 
This typically becomes the main prohibitive factor for cross 
sections at the level of $\sigma_{\mathrm{SI}} \gtrsim 10^{-28}\,\mathrm{cm}^2$~\cite{Emken:2018run}.

There have been a number of promising experimental proposals to probe these two open windows. Attempts 
to extend the sensitivity to DM-nucleus interactions into the sub-GeV realm include 
searches for Migdal electrons~\cite{Ibe:2017yqa,XENON:2019zpr} or bremsstrahlung photons~\cite{Kouvaris:2016afs}, 
accompanied by an intense low-threshold direct detection program in the development of novel detector 
concepts (for a recent review, see Ref.~\cite{Essig:2022dfa}).
Cross sections sufficiently large for DM to scatter inside the Earth before reaching underground detectors,
on the other hand,
can be probed by surface runs of conventional direct detection experiments (like the one performed by the CRESST 
collaboration~\cite{CRESST:2017ues}), or by targeting the expected diurnal modulation in the
signal in this case~\cite{Collar:1992qc,Collar:1993ss}. As far as this work is 
concerned, however, we will be interested in the role played by the irreducible astrophysical flux of
highly boosted DM that originates from cosmic ray collisions with DM particles in the Galactic halo (CRDM). 
This was pointed out only relatively recently~\cite{Bringmann:2018cvk,Cappiello:2018hsu}, and subverts the issue
of a loss in sensitivity by noting that a sub-dominant component of DM with velocities well above those in the Galactic halo can 
produce a detectable signal even if it is very light, i.e.~for DM masses (well) below $1 \, \mathrm{GeV}$. 
The sub-dominant nature of the flux 
naturally introduces a trade-off with the interaction rates that can be probed, quantitatively resulting in limits
at the level of $\sigma_\mathrm{SI} \sim 10^{-31}\,\mathrm{cm}^2$~\cite{Bringmann:2018cvk}.
Interestingly, CRDM does not only probe previously open parameter space at small DM masses but also results
in bounds extending into the relevant regime of the second open window described above. 
After this initial work pointed out the advantages of 
considering such a boosting mechanism, a large number of further analyses have addressed various aspects of the 
production~\cite{Alvey:2019zaa,DeRocco:2019jti,Dent:2019krz,Wang:2019jtk,Zhang:2020nis,Plestid:2020kdm,Su:2020zny,Cho:2020mnc,Guo:2020oum,Xia:2020apm,Dent:2020syp,Dent:2020qev,Emken:2021lgc,Das:2021lcr,Bell:2021xff,An:2021qdl,Feng:2021hyz,Wang:2021jic,Granelli:2022ysi,Darme:2022bew,Xia:2022tid,Bardhan:2022ywd}, 
attenuation~\cite{Bondarenko:2019vrb,McKeen:2022poo}, and 
detection~\cite{Ema:2018bih, Cappiello:2019qsw,Berger:2019ttc,Kim:2020ipj,Guo:2020drq,DeRoeck:2020ntj,Ge:2020yuf,Cao:2020bwd,Jho:2020sku,Lei:2020mii,Harnik:2020ugb,Ema:2020ulo,Bramante:2021dyx,Emken:2021vmf,PandaX-II:2021kai} 
of astrophysically boosted DM. 
For a recent comprehensive \mbox{(re-)analysis} of all of these aspects see, e.g.~Xia~{\it et al.}~\cite{Xia:2021vbz},
who stressed in particular that form-factor suppressed attenuation in the overburden seemingly allows us to exclude
cross sections much larger than $\sigma_\mathrm{SI} \sim 10^{-28}\,\mathrm{cm}^2$.

This article builds on this literature in three important ways: firstly, we point out that when DM acquires such large energies, 
inelastic scattering in the rock overburden above detectors such as Xenon-1T will at some point become the dominant
attenuation mechanism. 
As such, to avoid being over-optimistic in terms of how much parameter space is excluded, we show how 
to include this physical effect in a self-consistent manner and
derive the resulting bounds. Secondly, we broaden the applicability of these limits to models that are more realistic for DM 
with sub-GeV masses, moving beyond simplified contact interactions to
interactions mediated by vector or scalar mediators, or DM that has some internal structure. Finally, we argue that with 
these improvements, and when taking into account fully complementary constraints from cosmology,
there is generically no remaining open parameter space left unconstrained for nuclear cross sections exceeding 
$10^{-30}\,\mathrm{cm}^2$, 
for DM masses in the entire MeV to GeV range. We demonstrate that possible loopholes to this statement -- still allowing 
an open window at larger cross sections -- require a combination
of {\it (i)} questioning the principal ability of CRESST to probe DM masses down to the published limit of 
$m_\chi=140$\,MeV~\cite{CRESST:2017ues}  and 
{\it (ii)} choosing a rather narrow range of mediator masses $m_\phi\sim 30$\,MeV (or finite DM extent $r_\chi\sim10$\,fm).
For our numerical analysis throughout the article, we use the package \ds~\cite{Bringmann:2018lay}. The improved CRDM 
treatment reported in this work, including also updated cosmic ray fluxes and a more sophisticated use of form factors in the 
attenuation part, will be included in the next public release of the code.

\noindent The rest of the article is organized as follows: we start in section~\ref{sec:crdm} by briefly reviewing the production 
of CRDM and the attenuation of the subsequent flux on its way to the detector, establishing our notation and
setting up the basic formalism that our analysis relies on. In the next two sections, we discuss in more detail how to model nuclear 
form factors (section~\ref{sec:form_factors}) and the impact of inelastic scattering (section~\ref{sec:inel}) on the attenuation 
of the flux. In section~\ref{sec:m2}, we consider a number of 
generic options for the $Q^2$- and $s$-dependence of the scattering amplitude that are more realistic than assuming a constant 
cross section. We complement this in section~\ref{sec:hexaquark} with the analysis of a specific example, namely a
baryonic DM candidate that has been argued to evade traditional direct detection bounds despite its relatively
strong interactions with nuclei. We conclude and summarise our results in section~\ref{sec:conclusions}.

\section{Cosmic-ray upscattering of dark matter}
\label{sec:crdm}

We describe here, in turn, how initially non-relativistic DM particles in the Galactic halo are up-scattered by cosmic rays (CRs),
how the flux of these relativistic CRDM particles is attenuated before reaching detectors at Earth, and
how to compute the resulting elastic scattering rate in direct detection experiments.

\medskip
\noindent\textbf{Production:} The basic mechanism that we consider is the elastic scattering of CR nuclei $N$,  
with a flux of ${{d\Phi_N}}/{dT_N}$, 
on non-relativistic DM particles $\chi$ in the Galactic halo. For a DM mass $m_\chi$ and 
density profile $\rho_\chi(\mathbf{r})$, this induces a relativistic CRDM flux incident on Earth 
of~\cite{Bringmann:2018cvk,Bondarenko:2019vrb} 
\bea
\frac{d\Phi_{\chi}}{dT_\chi}&=&\int\frac{d\Omega}{4\pi} \int_{\rm l.o.s.} \!\!\!\!\!\!d\ell \, \frac{\rho_\chi}{m_\chi} 
\sum_N
\int_{T_N^\mathrm{min}}^\infty d T_N\, \frac{d \sigma_{\chi N} }{dT_\chi} \frac{{d\Phi_N}}{dT_N}\\
&\equiv&  
D_\mathrm{eff} \frac{\rho_\chi^\mathrm{local}}{m_\chi}  
\sum_N
\int_{T_N^\mathrm{min}}^\infty d T_N\, \frac{d \sigma_{\chi N} }{dT_\chi} \frac{{d\Phi^\mathrm{LIS}_N}}{dT_N}
\label{eq:chiCR}
\,.
\eea
Here $\mathbf{r}$ denotes the Galactic position, and 
${d \sigma_{\chi N} }/{dT_\chi}$ is the differential elastic scattering cross section
for accelerating a DM particle to a kinetic recoil energy $T_\chi$. 
For DM particles initially at rest, this requires a minimal CR energy $T_N^\mathrm{min}$ of
\be
\label{eq:Tmin}
T_N^\mathrm{min}=
\left\{
\begin{array}{ll}
\left( \frac{T_\chi}{2}  - m_N\right) \left[
1-\sqrt{1+\frac{2 T_\chi}{m_\chi}\frac{\left(m_N + m_\chi\right)^2}{\left(2m_N - {T_\chi}\right)^{2}}}
\right] & \quad\mathrm{for~}T_\chi<2m_N\\
\sqrt{\frac{m_N}{m_\chi}} \left(m_N + m_\chi\right) & \quad\mathrm{for~}T_\chi=2m_N\\
\left( \frac{T_\chi}{2}  - m_N\right) \left[
1+\sqrt{1+\frac{2 T_\chi}{m_\chi}\frac{\left(m_N + m_\chi\right)^2}{\left(2m_N - {T_\chi}\right)^{2}}}
\right] & \quad\mathrm{for~}T_\chi>2m_N
\end{array}
\right. \,.
\ee
Furthermore, in the second line of Eq.~(\ref{eq:chiCR}), we have introduced an effective distance $D_{\rm eff}$ that allows us to 
express the CRDM flux in the solar system in terms of the relatively well measured {\it local} interstellar CR flux, 
${{d\Phi_N}^{\rm LIS}}/{dT_N}$, and the {\it local} DM density, for which we adopt 
$\rho^\mathrm{local}_\chi=0.3\,\mathrm{GeV}/\mathrm{cm}^3$~\cite{Read:2014qva} (noting that our final limits are 
independent of this choice). The advantage of this parameterisation is that uncertainties deriving from the integration 
over the volume relevant for CRDM production, $\int d\Omega \int \!d\ell$,  are captured in a single 
phenomenological parameter $D_\mathrm{eff}$. Indeed, despite the complicated underlying physics, this parameter is 
surprisingly well constrained,
with uncertainties dominated by the vertical extent of the confinement zone of Galactic CRs. 
In what follows, we will use a fiducial value of $D_{\rm eff}=10$\,kpc.\footnote{%
When assuming  an Einasto profile~\cite{Einasto:2009zd} for the DM density, and a cylindric CR diffusion model 
tuned with {\sf GalProp}~\cite{Strong:1998pw} to describe the observed flux of light CR nuclei,   
a more detailed analysis reveals that $D_{\rm eff}$ varies between $\sim9$\,kpc and 
$\sim11$\,kpc for DM recoil energies above 1\,MeV~\cite{Xia:2021vbz} . 
}
We note that our final limits only depend logarithmically on this quantity, for large interaction rates,
or scale as $D_{\rm eff}^{-1/2}$ when attenuation in the soil or atmosphere is inefficient, respectively.

When computing the CRDM flux in Eq.~(\ref{eq:chiCR}), we take into account the four most abundant
CR species, $N=\{p,{\rm He},{\rm C}, {\rm O}\}$, for which high-quality determinations of the local 
interstellar fluxes exist~\cite{Boschini:2018baj}.  The fluxes of heavier nuclei are subject to significant 
uncertainties for the energies of interest to us, see e.g.~the discussion in Ref.~\cite{Boschini:2020jty}, not least due to 
apparent discrepancies between AMS-02 data~\cite{AMS:2018tbl,AMS:2018cen,AMS:2020cai} and 
earlier measurements. We also note that the CRDM flux contribution from these heavier elements is 
strongly form-factor suppressed at large $T_\chi$, see section \ref{sec:form_factors}, and hence 
anyway not relevant for constraining DM with masses $m_\chi\gtrsim0.1$\,GeV.

\medskip
\noindent\textbf{Attenuation:} On its way to the detector, the CRDM flux given by Eq.~(\ref{eq:chiCR}) is attenuated due to 
scattering of the CRDM particles with nuclei in the atmosphere and soil (overburden) above the experimental
location. This effect can be well modelled by the energy loss equation
\be
\label{eq:eloss}
\frac{dT_\chi^z}{dz}=-\sum_N n_N\int_0^{\omega_\chi^\mathrm{max}}\!\!\!d\omega_\chi\,\frac{d \sigma_{\chi N}}{d\omega_\chi} \omega_\chi\,,
\ee
which can be used to relate the average kinetic energy at depth $z$, $T_\chi^z$, to an initial energy 
$T_\chi$ at the top of the atmosphere. 
Here, the sum runs over the nuclei $N$ in the overburden,
i.e.~no longer over the CR species, and $\omega_\chi$ is the {\it energy loss} of a DM particle
in a single collision. 
For elastic scattering, $\omega_\chi$ is equal to the nuclear recoil energy $T_N$.
In that case, the maximal energy loss of a DM particle with initial kinetic energy $T_\chi^z$
is given by
\be
\label{eq:tmax}
\omega_\chi^\mathrm{max}=T_N^\mathrm{max}=\frac{2m_N}{s}\left[\left(T_\chi^z\right)^2+2m_\chi T_\chi^z\right],
\ee 
where 
\be
\label{eq:sdef}
s=(m_N+m_\chi)^2+2m_N T_\chi^z
\ee
is the (squared) CMS energy of the process. For inelastic scattering on the other hand, which we will discuss in more detail 
in section \ref{sec:inel}, the energy loss can in principle be as high as $\omega_\chi^\mathrm{max}=T_\chi^z$.
For the purpose of this work we will mostly be interested in the Xenon-1T
detector, located at a depth of $z=1.4\, \text{km}$  in the Gran Sasso laboratory. In this case the 
limestone overburden has a density of 2.71 g/cm$^3$~\cite{Miramonti:2005xq},
mostly consisting of an admixture of CaCO$_3$ and MgCO$_3$, and attenuation in the
atmosphere can be neglected; in terms of weight percentages
the dominant elements are O (47.91\%), Ca (30.29\%), C (11.88\%), Mg (5.58\%), Si (1.27\%),
Al (1.03\%) and K (1.03\%)~\cite{Wulandari:2003cr}. We note that Eq.~(\ref{eq:eloss}) only provides an approximate 
description of the stopping 
effect of the overburden, which is nonetheless sufficiently accurate for our purposes. For a detailed comparison of this 
approach with Monte Carlo simulations of individual particle trajectories, see 
Refs.~\cite{Emken:2017qmp,Emken:2018run,Mahdawi:2018euy,Emken:2019hgy,Xia:2021vbz}

\medskip
\noindent\textbf{Detection:} The elastic scattering rate of relativistic CRDM particles arriving at underground detectors 
like the Xenon-1T experiment is given by
\be
\label{eq:gammarate}
 {\frac{d\Gamma_N}{d T_{N}}=
 \int_{T_\chi^{\rm min}}^\infty \!\!dT_\chi\ 
 \frac{d \sigma_{\chi N}}{dT_N} \frac{d\Phi_\chi}{dT_\chi}} \,.
\ee
Note that the above integral is over the energy of the DM particles \emph{before} entering the atmosphere. 
On the other hand, the elastic scattering cross section ${d \sigma_{\chi N}}/{dT_N} $ must still be evaluated at the actual 
DM energy, $T_\chi^z$, at the detector location, which requires numerically solving Eq.~(\ref{eq:eloss}) 
for $T_\chi^z(T_\chi)$. The lower bound on the integral then represents the minimal {\it initial} CRDM energy 
that is needed to induce a nuclear recoil of energy $T_N$ {\it at depth $z$}, i.e.
$T_\chi^{\rm min}=T_\chi(T_\chi^{z, \mathrm{min}})$. This can be obtained by inverting the solution of Eq.~(\ref{eq:eloss}),
where $T_\chi^{z, \mathrm{min}}$ is given by the right-hand side of Eq.~(\ref{eq:Tmin}) under the replacement
$(T_\chi,m_\chi,m_N)\to(T_N,m_N,m_\chi)$.
In general, the elastic nuclear scattering cross section 
${d \sigma_{\chi N}}/{dT_N} $ 
is a function of both $s$ and the (spatial) momentum transfer, 
\be
\label{eq:q2}
Q^2=2m_N T_N\,.
\ee
If the dependence on $s$ can be neglected or the (dominant) dependence on $Q^2$ factorizes -- as in the case of
standard form factors -- then the rate in the detector given in Eq.~(\ref{eq:gammarate}) will have an {\it identical}
$Q^2$-dependence as compared to the corresponding rate expected from the standard population of 
non-relativistic halo DM. As pointed out in Ref.~\cite{Bringmann:2018cvk}, this salient feature makes it possible to 
directly re-interpret published limits on the 
latter (conventionally expressed as limits on the scattering cross section with protons) into limits on the 
former. Otherwise, for an accurate determination of the expected count rate in 
a given analysis window, one would in principle have to also model the detector response in the 
evaluation of Eq.~(\ref{eq:gammarate}) and then infer limits based on the full detector likelihood 
(e.g.~with a tool like {\sf DDCalc}~\cite{GAMBITDarkMatterWorkgroup:2017fax,GAMBIT:2018eea}).

\section{Nuclear form factors}
\label{sec:form_factors}

The target nuclei used in direct detection experiments 
are typically larger than the de Broglie wavelength of DM with standard Galactic velocities, 
at least for heavy nuclei, implying that the incoming DM particles only `see' part of the nucleus. 
Since the elastic scattering process is fundamentally induced by a coupling between DM and the 
constituents of these nuclei, this means that it should be suppressed by a 
nuclear form factor, $G^2(Q^2)$, compared to the naive expectation that the nuclear cross section 
is merely a coherent sum of the cross sections of all the constituents (for recent pedagogic 
accounts of conventional direct DM searches, see e.g.~Refs.~\cite{DelNobile:2021icc, Cooley:2021rws}).\footnote{%
We focus here on spin-independent elastic scattering.  For {\it spin-dependent} scattering,  
the sum would not be coherent and hence generally result in much smaller cross sections.
This prevents standard DM from being stopped in the overburden before reaching the experimental
location -- unless the scattering cross section {\it per nucleon} is so large that it becomes incompatible with other
astrophysical constraints. 
A detailed treatment of attenuation in the Earth's crust is, hence, less relevant in this case. 
}
For CRDM, this effect is amplified, given the smaller de Broglie wavelengths
associated to the faster moving upscattered DM particles. 

These nuclear form factors are essentially Fourier transforms of the number density of nucleons inside
the nucleus, usually approximated by the experimentally easier accessible charge density. A common
parameterization is the one suggested by Helm~\cite{Helm:1956zz}, which is based on modelling 
the nucleus as a hard sphere with a Gaussian smearing (in configuration space).  For heavy nuclei we follow instead a slightly
more accurate approach and implement model-independent form factors~\cite{Duda:2006uk}, based
on elastic electron scattering data. Concretely, we implement their Fourier-Bessel (FB) expansion approach,
with parameters taken from Ref.~\cite{DeVries:1987atn}. For nuclei where the FB parameters
are not available, notably Mg and K, we use model-independent Sum of Gaussians (SOG) form factors instead. 

For $Q^2\gg (0.1\,\mathrm{GeV})^2$ one starts
to resolve the inner structure of the nucleons themselves, which we discuss in more detail in section \ref{sec:inel}. 
Let us however briefly mention 
that in the case of He, this effect is already largely captured by the above description in that we take the 
SOG form factors from Ref.~\cite{DeVries:1987atn} (thus improving on the simple dipole prescription used, e.g.,
in Ref.~\cite{Bringmann:2018cvk}). For the proton, we adopt the usual dipole {\it nucleon} form factor,
noting that the {\it nuclear} form factor would formally equal unity,
\be
\label{eq:Gp}
G_p^2(Q^2)=\left(1+ Q^2/\Lambda_p^2 \right)^{-4}\,,
\ee
with $\Lambda_p=0.843$\,GeV. This provides a very good fit to experimental data up to momentum
transfers of at least $Q^2\sim1$\,GeV$^2$, with an agreement of better than 10\% for 
$Q^2\leq10$\,GeV$^2$~\cite{Perdrisat:2006hj,Punjabi:2015bba}. 
We note that our final results are highly insensitive to such large momenta.

In the rest of the section, we will briefly describe the impact of nuclear form factors on 
the CRDM flux and the attenuation of this flux on its way to the detector.
In both cases the effect is sizeable, motivating the need for a precise modelling of $G^2(Q^2)$.

\subsection{Impact on production}
\label{sec:FF_prod}

\begin{figure}[t]
\begin{center}
\includegraphics[width=0.99\textwidth]{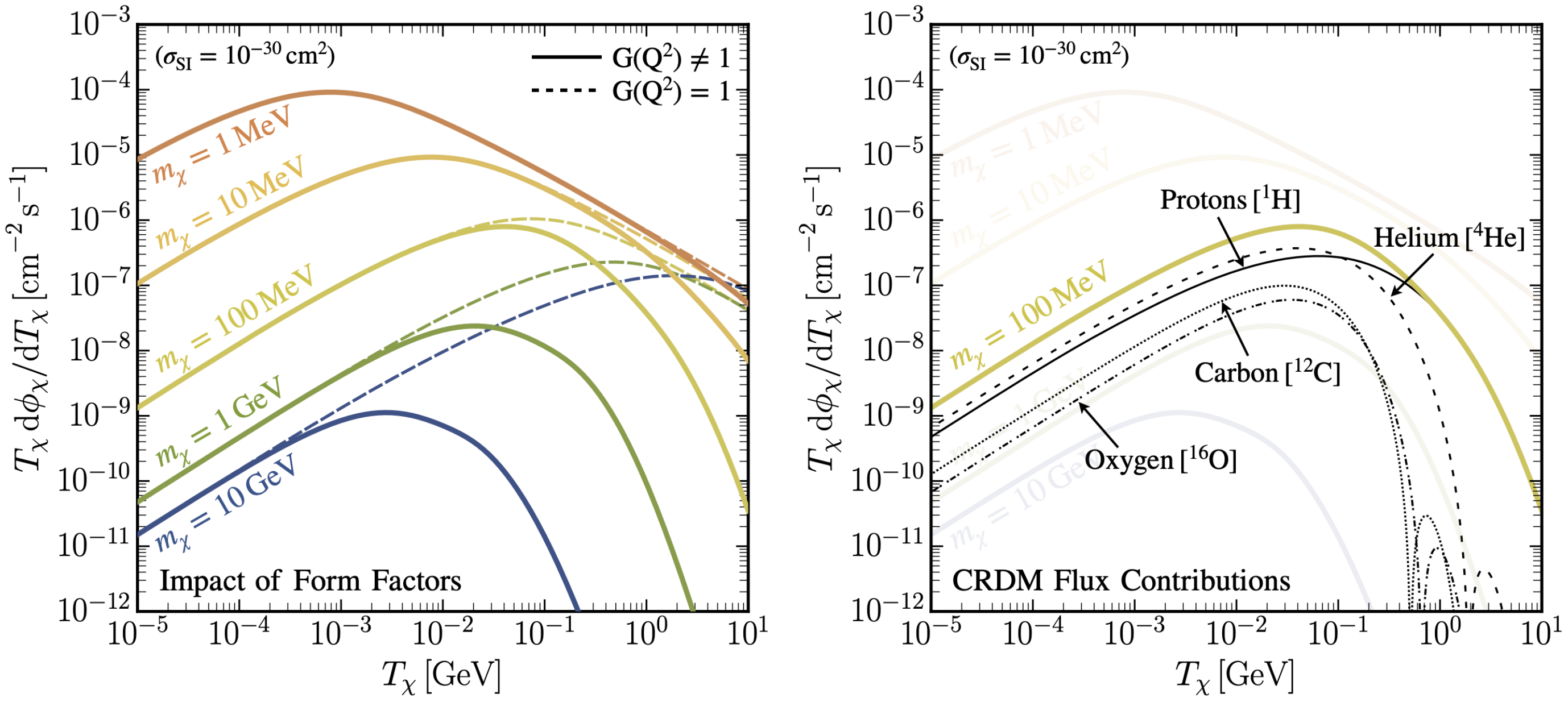}
\caption{{\it Left panel.} Expected CRDM fluxes for DM masses $m_\chi=0.001, 0.01, 0.1,1,10$\,GeV, 
from top to bottom, assuming a constant spin-independent scattering cross section of 
$\sigma_{\rm SI}^{p,n}=10^{-30}\,\mathrm{cm}^2$ (solid lines). The effect of inelastic scattering is neglected.
Dashed lines show the CRDM fluxes that would result when not taking into account 
the effect of form factors. 
{\it Right panel.} Black lines indicate the individual contributions to the CRDM flux from scattering on CR $p$, He, C and O,
for the example of $m_\chi=100$\,MeV. Other lines (highlighted only for the  $m_\chi=100$\,MeV case) show the
total flux, as in the left panel.
}
\label{fig:flux_ff}
\end{center}
\end{figure}

The solid lines in Fig.~\ref{fig:flux_ff} show the expected CRDM flux before attenuation, cf.~Eq.~(\ref{eq:chiCR}),
for a range of DM masses. For the purpose of this figure, we have assumed a constant elastic 
scattering cross section $\sigma_{\rm SI}^p=\sigma_{\rm SI}^n$ on nucleons, i.e.~a nuclear cross section
given by
\be
\frac{d \sigma_{\chi N}}{dT_\chi} = \mathcal{C}^2
\times \frac{\sigma_{\rm SI}^p}{T_\chi^{\rm max}} \times G^2(2T_\chi m_\chi)\,.
\label{eq:siconst}
\ee
Here, 
\be
\label{eq:c_coh}
\mathcal{C}^2= A^2\frac{\mu_{\chi N}^2}{\mu_{\chi p}^2}
\ee
describes the usual coherent enhancement, in this case proportional to the square of the atomic number $A$ 
of nucleus $N$. In the rest of the expression, $\mu_{\chi N}$ ($\mu_{\chi p}$) is the reduced mass of the 
DM/nucleus (DM/nucleon) system and
the maximal DM energy $T_\chi^{\rm max}$ that can result from a CR nucleus with energy $T_N$ 
is given by the right-hand side of Eq.~(\ref{eq:tmax}) after replacing $T_\chi^z \to T_N$ and $m_\chi\leftrightarrow m_N$.

In the left panel of the figure, we show that neglecting nuclear form factors (dashed lines) would lead to a 
significant overestimate of the CRDM flux at high energies. For $m_\chi\gtrsim0.1$\,GeV, the form factor
suppression even becomes the dominant effect to determine the overall normalization of the flux,
while for lower DM masses, the peak of the distribution is entirely determined by the fact that the 
CR flux itself peaks at GeV energies. This suppression in the flux leads to a rapid deterioration 
of CRDM limits.
Modelling form factors correctly is thus particularly important for the highest DM masses that can be
probed by cosmic-ray upscattering, i.e.~for $m_\chi \sim 1 - 10 \, \mathrm{GeV}$.

In the right panel of Fig.~\ref{fig:flux_ff}, the contributions from the individual CR nuclei to the  
CRDM flux are shown. At low energies the dominant contribution is always from Helium, closely followed by the one from protons. 
The high-energy part of the CRDM flux, on the other hand, 
is almost exclusively due to CR protons because the contribution from heavier CR nuclei is 
heavily form-factor suppressed. In addition, for $m_\chi\gtrsim1$\,GeV, the 
peak amplitude of the CRDM flux -- which typically has the most constraining
power in direct detection experiments -- is almost exclusively  determined by CR $p$ and He nuclei
 (see also Fig.~\ref{fig:attenuation_ff} below to better gauge the relevant range of energies {\it after} attenuation
 in the overburden).
For lower DM masses, on the other hand, including further high-$Z$ CR species than those taken into account 
here could in principle increase the relevant part of the CRDM flux by up to $\sim50$\,\%~\cite{Xia:2021vbz}. 
In what follows, we conservatively neglect these contributions, in view of both the larger uncertainties in the underlying
CR fluxes and the fact that we are mainly interested in DM masses around the GeV scale.

\subsection{Impact on attenuation}
\label{sec:FF_att}

\begin{figure}[t]
\begin{center}
\includegraphics[width=0.8\textwidth]{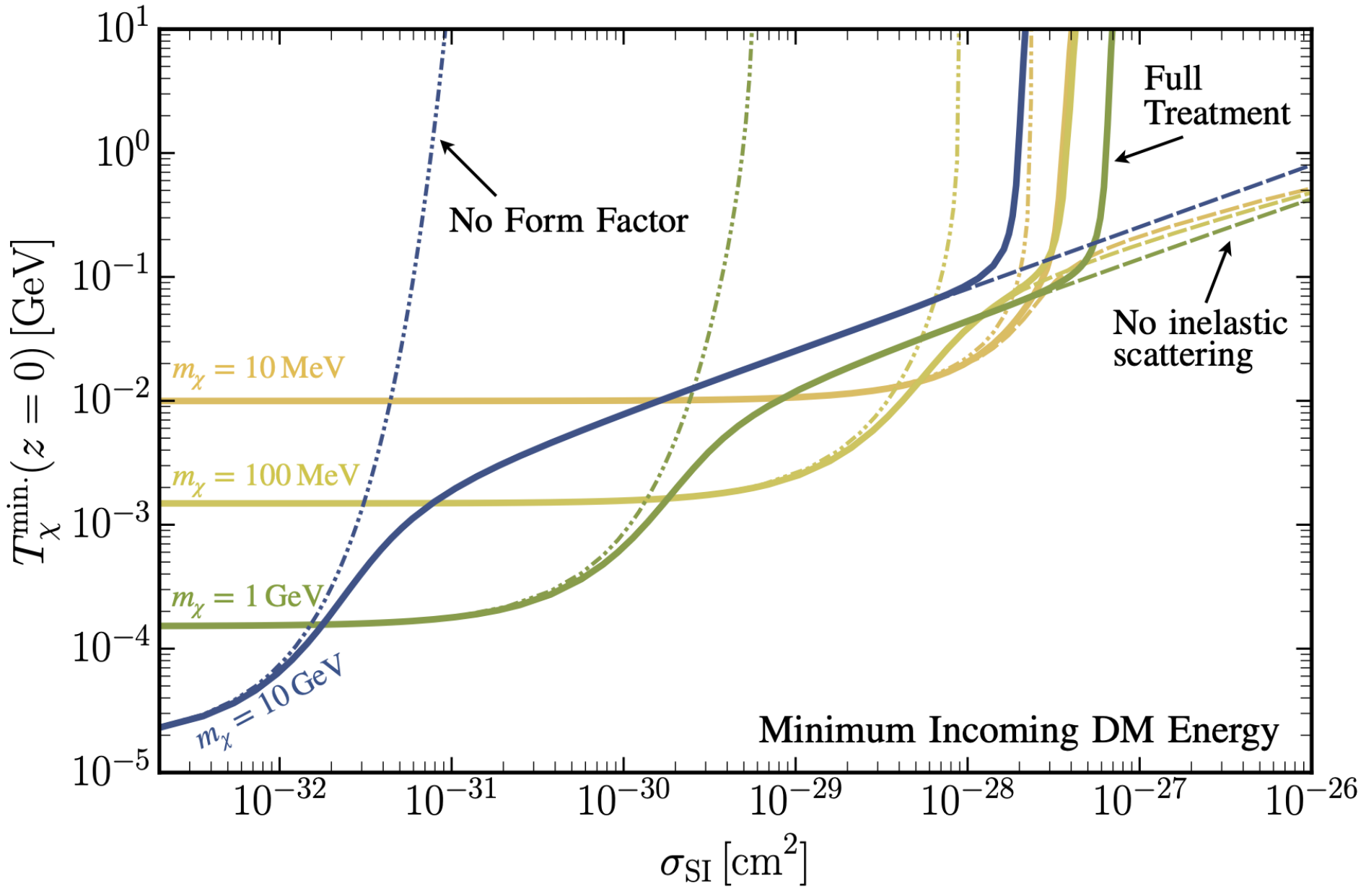}
\caption{%
Minimal kinetic energy $T_\chi$ that a DM particle must have at the surface of the Earth ($z=0$) in order 
to trigger a signal in the Xenon-1T experiment, as a function
of a (constant) spin-independent scattering cross section $\sigma_{\rm SI}^{p,n}$ on nucleons.
 Different colors correspond to different DM masses, 
 as  in Fig.~\ref{fig:flux_ff}.
 Dash-dotted lines show the kinetic energies that would be necessary when computing the attenuation in the 
zero momentum transfer limit. Dashed lines illustrate the effect of adding
the expected form factor suppression, cf.~section \ref{sec:form_factors}, while solid 
lines show the result of our full treatment, including also inelastic scattering events 
(discussed in section \ref{sec:inel}).
}
\label{fig:attenuation_ff}
\end{center}
\end{figure}

We now turn our attention to assessing the effect that the form factor suppression has on the attenuation of DM
particles on their way to the detector in a direct detection experiment. For concreteness we will again focus
on the case of Xenon-1T, where Xe nuclei recoiling with an energy of at least
$T_{\rm Xe}=4.9$\,keV trigger a detectable signal~\cite{XENON:2018voc}. In 
Fig.~\ref{fig:attenuation_ff}, we show the minimal initial DM energy that is required to kinematically allow
for this, after penetrating through the Gran Sasso rock. In practice this is done by numerically solving Eq.~(\ref{eq:eloss}) with
\ds. Dash-dotted lines indicate the result when conservatively
assuming that the stopping power in the overburden is as efficient as in the zero-momentum transfer
limit (as in Ref.~\cite{Bringmann:2018cvk}), while dashed lines show the effect of adding the additional
form factor suppression for high $Q^2$ (as in Refs.~\cite{Bell:2021xff,Xia:2021vbz}). 
Solid lines, finally, demonstrate the effect of also adding the attenuation power of inelastic scattering events,
as described in detail below in Section \ref{sec:inel}.

For small cross sections, attenuation is inefficient and, as expected, the three approaches give the 
same answer. In this limit, the difference in the required DM energy is entirely due to the well-known
kinematic effect, cf.~Eq.~(\ref{eq:Tmin}), that lighter particles require a higher energy to induce a 
given recoil of much heavier particles
(up to a minimum energy of $T_\chi\geq\sqrt{m_{\rm Xe}T_{\rm Xe}/2} =17.3$\,MeV in the limiting case where $m_\chi\to0$).
Correspondingly, this also means that the CRDM fluxes cannot actually be probed by Xenon-1T 
for the entire range of $T_\chi$ shown in Fig.~\ref{fig:flux_ff};
unless $m_\chi\lesssim10$\,MeV, however, the lowest detectable energy is always smaller
 than the energy at which the CRDM flux peaks.
 
For large cross sections, on the other hand, Fig.~\ref{fig:attenuation_ff} shows a 
pronounced difference between the three 
approaches: while in the case of a constant cross section (dash-dotted lines) the energy loss equation
results in an exponential attenuation, adding form factors (dashed lines) implies that the required initial 
DM energy only rises as the square root of the scattering cross section in the $Q^2=0$ limit.
In fact, we note that this is exactly the behaviour one would expect from Eq.~(\ref{eq:eloss}) for a 
cross section that falls off very rapidly at large momentum transfers. 
Comparing again to Fig.~\ref{fig:flux_ff},
this correspondingly enlarged range of kinetic energies that becomes kinematically accessible to Xenon-1T will 
inevitably lead to significantly larger rates in the detector -- which, indeed, is exactly the conclusion reached in
Refs.~\cite{Bell:2021xff,Xia:2021vbz}. However, such a strong 
suppression of the physical stopping power of the Gran Sasso rock for a relativistic particle is highly 
unphysical. As we discuss in the next section, this is simply because the DM particles will start to scatter off the constituent 
nucleons themselves, albeit not 
coherently across the whole nucleus. Adding this effect (solid lines), 
results again in exponential attenuation in the overburden -- though only at significantly larger cross sections 
than what would be expected when adopting a constant cross section for simplicity.

\section{Inelastic Scattering}
\label{sec:inel}

Our discussion so far has largely neglected the impact of inelastic scattering events of relativistic DM particles incident on nuclei 
at rest, or {\it vice versa}. Physically, the inclusion of inelastic scattering processes is non-negotiable and should be considered in 
a full treatment. This is because, whilst the form factor suppression described above is the relevant feature in the transition from 
coherently scattering off the whole nucleus to only parts of it, once the DM or nucleus transfers a sufficiently large amount 
of energy $\omega$, the scattering will probe individual nucleon-, or even quark-level processes. The result is an additional 
contribution to the total scattering cross section that can easily dominate 
in the large energy transfer regime. 
As far as CRDM limits are concerned, the most important effect 
that the inclusion of inelastic scattering modifies is the attenuation of the flux through the Earth or atmosphere.
Not including it, therefore, will lead to an overly optimistic estimate as to the amount of parameter space that is ruled out via this 
mechanism.\footnote{In order to keep our results conservative, we neglect the effect of inelastic scattering on CRDM {\it 
production} in our analysis. We leave the study of this additional contribution of the flux to future work, noting that we 
expect it to mostly improve limits for larger DM masses (where the form factor suppression nominally leads 
to a significant reduction of the CRDM flux, see Fig.~\ref{fig:flux_ff}).}
Let us note that inelastic scattering of {\it non-relativistic} DM, resulting in the excitation of low-lying states in the target nuclei,  
was previously both studied 
theoretically~\cite{Baudis:2013bba, McCabe:2015eia, Kouvaris:2016afs, Hoferichter:2018acd}
and searched for experimentally~\cite{XMASS-I:2014lnb,XENON:2017kwv,Lehnert:2019tuw,XENON:2020fgj}. 
Here we concentrate on different types of inelastic processes that are only accessible to nuclei scattering off 
high-energy DM particles.

The rest of this section is organised as follows: firstly we give a qualitative description of the most important 
inelastic scattering processes, such as the excitation of hadronic resonances or quasi-elastic scattering 
off individual nucleons. Secondly, we explain how we obtain a quantitative estimate of these complicated nuclear interactions 
by making a direct analogy to the case of neutrino-nucleus scattering. In this regard, we make use of the public code 
\texttt{GiBUU}~\cite{Buss:2011mx,gibuuweb}.
Finally, we will explain how to build this into the formalism described in section~\ref{sec:crdm} in terms of the DM energy loss, see 
Eq.~\eqref{eq:eloss}.

\subsection{Scattering processes and associated energy scales}
There are a number of relevant contributions to scattering cross sections on nuclei that are associated to certain 
characteristic energies or nuclear length scales. In the highly non-relativistic limit, as described above,
coherently enhanced elastic scattering dominates. At somewhat higher energies, more specifically momentum transfers
corresponding to (inverse) length scales smaller than the size of the nucleus,
the elastic scattering becomes form factor suppressed -- a description which physically assumes a smooth distribution of 
scattering centres throughout the nucleus. The main characteristic of elastic scattering in both of these regimes is that 
the energy loss of the incident DM particle is uniquely related to the momentum transfer by $\omega=Q^2/(2m_N)$.

This relation no longer holds for inelastic scattering processes, which are expected to become relevant at even higher energies. 
For our purposes, these inelastic processes can be broadly split up into three scattering regimes, depending
on the energy that is transferred (see also Fig.~\ref{fig:mchi_1GeV} below, as well as a review~\cite{Formaggio:2012cpf} 
for the discussion of the analogous situation in the case of neutrino-nucleus scattering):
       
\begin{itemize}
      \item \textbf{Quasi-Elastic Scattering} \textbf{(}$\mathbf{\omega \gtrsim 10^{-2}}$\,\textbf{GeV):} At suitably large 
	energy transfers, the form factor suppression cannot be totally physical. This is because  
	the incident DM particles will probe directly the constituent nucleons, which are 
	inherently not smoothly distributed. \emph{Quasi-elastic scattering} (QE) dominates for 
	$10^{-2}\,\mathrm{GeV} \lesssim \omega \lesssim 1 \, \mathrm{GeV}$, and describes this situation, 
	i.e.~where the dominant scattering is directly off {\it individual} protons (and neutrons) inside the nucleus, 
	$\chi\, p (n) \rightarrow \chi\, p (n)$.

	\item \textbf{Excitation of Hadronic Resonances} \textbf{(}$\mathbf{\omega \gtrsim 0.2}$\,\textbf{GeV):} At higher energies 
	still, DM-nucleon scattering can excite nuclear resonances such as 
	$\chi \, p \rightarrow \chi \, (\Delta \rightarrow p \pi^0)$ etc., leading to a wide variety of hadronic final states. Often, the contribution due to the lowest lying 
	$\Delta$ resonances (DR) is distinguished from contributions from higher resonances (HR) since the former can  
	be well resolved and starts playing role at considerably smaller transferred energies. 
	In a complicated nucleus such as ${^{16}}\mathrm{O}$, both the QE and resonance contributions to the scattering cross section must be resolved numerically, 
	taking into account effects such as the nuclear potential and spin statistics. 

	\item \textbf{Deep Inelastic Scattering} \textbf{(}$\mathbf{\omega \gtrsim 1}$ \textbf{GeV):} Most DM couplings to 
	nuclei and nucleons result from more fundamental couplings to quarks or gluons. 
	As such, once the energy transfer is large enough to probe the inner structure of the nucleons 
	($\omega \gtrsim 1 \, \mathrm{GeV}$), then \emph{deep inelastic scattering} (DIS) of DM with partons inside the nucleons
	can occur. Again, this should be 
	resolved numerically to give an accurate estimate of the impact at the level of the scattering cross section.
\end{itemize}

\subsection{Computation of the inelastic cross section for neutrinos}
\label{sec_gibuu}

Due to the complicated nuclear structure of the relevant atomic targets in the Earth, 
or in the composition of cosmic rays, 
it is typically not possible to analytically compute all the contributions to DM-nucleus scattering 
described above. Instead, to estimate their impact on our 
conclusions and limits, we will make a direct connection with the physics of neutrino-nucleus scattering for which numerical codes 
-- such as \texttt{GiBUU}~\cite{Buss:2011mx} -- are capable of generating the relevant differential cross sections.

In more detail, we draw the analogy between neutral current neutrino-nucleon scattering via processes such as 
$\nu \, p \rightarrow \nu \, p$ and DM-nucleon scattering. Numerically modelling the neutral current quasi-elastic scattering, 
resonances and deep inelastic scattering as a function of the energy transferred to the nucleus, $\omega$, allows us to 
understand the relative importance of these processes as a function of the incoming neutrino energy
(or DM kinetic energy $T_\chi$). Of course, since 
these codes are tuned for neutrino physics, simply outputting the differential cross sections such as 
$\mathrm{d}\sigma_{\nu N} / \mathrm{d}\omega$ is not sufficient. To map the results onto DM, see 
section \ref{sec:map_dm} below for further details, we should re-scale the results so 
as to respect both the relative interaction strengths and model dependences such as e.g. the mediator mass. In general, we 
expect this approach to provide a good estimate of the DM-nucleus cross section (at least) for contact interactions and  
scattering processes dominated by mediators in the $t$-channel.

At the level of implementation, we choose the settings in the \texttt{GiBUU} code described in Tab.~\ref{tab:gibuu} (see end of text). 
Since we are interested in quantifying the effect of inelastic scattering on the attenuation of the CRDM flux as it passes 
through the Earth, 
we mostly focus on the total inelastic scattering cross section, i.e.~the sum over all the processes described in the 
previous section. We numerically calculate this for the most abundant nuclei in the Gran Sasso rock, 
$N = \{\mathrm{O}, \mathrm{Ca}, \mathrm{C}, \mathrm{Mg}, \mathrm{Si}, \mathrm{Al}, \mathrm{K}\}$.
Fundamentally, inelastic cross sections are expressed in terms of double-differential cross sections 
like $\mathrm{d}^2 \sigma_{\nu N} / \mathrm{d}Q^2 \mathrm{d}\omega$, since for inelastic scattering $Q^2$ and $\omega$ are 
independent variables. 
For integrating the energy loss equation, Eq.~\eqref{eq:eloss}, however, it suffices to compute
\begin{equation}
\frac{\mathrm{d} \sigma_{\nu N}}{ \mathrm{d}\omega} \equiv \int_{Q^2} \frac{\mathrm{d}^2 \sigma_{\nu N}}{ \mathrm{d}Q^2 \,\mathrm{d}\omega} \,\mathrm{d}Q^2\,.
\end{equation}
On the other hand, the full information about the $Q^2$-dependence of 
$\mathrm{d}^2 \sigma_{\nu N} / \mathrm{d}Q^2 \mathrm{d}\omega$ provided by \texttt{GiBUU} still remains a highly
useful input to our analysis. This is because  the double-differential cross 
sections of the individual inelastic processes turn out to sharply peak at values of $Q^2$ that have simple relations to $\omega$.
For example, the peak position for the QE contribution corresponds to the `elastic' relation~\eqref{eq:q2} for nucleons. 
As described below, this information will be used 
for setting realistic reference values of $Q^2$ to capture the model-dependence of the DM cross sections.

\subsection{Mapping to the dark matter case}
\label{sec:map_dm}

Having described the technical details of how we obtain the neutrino-nucleus inelastic cross sections using \texttt{GiBUU}, we 
now turn our attention to the mapping of these quantities onto DM models. 
This is a necessary step for two broad reasons: \emph{(a)} the interaction strength governing the DM-nucleus interactions is 
typically very different from the neutrino-nucleus SM value, and \emph{(b)} the way the interaction proceeds via e.g.~a contact 
interaction or mediator exchange can lead to substantially different kinematics and non-trivial $Q^2$- or $s$-dependences.

The total scattering cross section $\mathrm{d}\sigma_{\chi N}/\mathrm{d}\omega$ consists of the coherent elastic scattering contribution that we compute analytically for each of the models considered in this work, and the inelastic scattering cross section that we want to estimate based on the \texttt{GiBUU} output:
\begin{align}\nonumber
\frac{\mathrm{d}\sigma_{\chi N} }{ \mathrm{d}\omega} &= \left.\frac{\mathrm{d}\sigma_{\chi N} }{ \mathrm{d}\omega}\right|_{\mathrm{el}}+\left.\frac{\mathrm{d}\sigma_{\chi N} }{ \mathrm{d}\omega}\right|_{\mathrm{inel}}\\ \label{eq:rescaling}
&\equiv  
	\left.\frac{\mathrm{d}\sigma_{\chi N} }{ \mathrm{d}\omega}\right|_{\mathrm{el}, Q^2=2\omega m_N} + \sum_{i} \left.\frac{\mathrm{d}\sigma_{\mathrm{SI}} }{ \mathrm{d}\omega}\right|_{\mathrm{el}, Q^2=Q_{i,\mathrm{ref}}^2} 
\times I_{\chi,i}(T_\chi,\omega)\,.
\end{align}
Here $\left.\mathrm{d}\sigma_{\mathrm{SI}} / \mathrm{d}\omega\right|_{\mathrm{el}}$ is the differential
DM-nucleon elastic cross section, excluding nucleon form factors such as the one given in 
Eq.~(\ref{eq:Gp}). 
The sum runs over the various individual processes, $i\in$(QE, DR, HR, DIS),
which all have characteristic reference values of $Q^2=Q^2_{i, \mathrm{ref}}(\omega)$ where
the respective inelastic cross section peaks. In the second step above, we thus choose to rescale the inelastic scattering 
events to the elastic scattering off a point-like nucleon. 
This rescaling is motivated by the fact that for inelastic contributions like QE, the underlying process is much better 
described by scattering on individual nucleons than on the entire nucleus.
The factor
\be
I_{\chi,i}(T_\chi,\omega) \equiv \frac{\mathrm{d}\sigma^i_{\chi N} /\mathrm{d}\omega \big|_{\mathrm{inel}}}
	{\mathrm{d}{\sigma}_{\mathrm{SI}} /\mathrm{d}\omega \big|_{\mathrm{el},Q^2=Q^2_{i, \mathrm{ref}}}}
\ee
thus quantifies the ratio of the inelastic scattering process on a nucleus to the elastic scattering on an individual
nucleon.

We now make the simplifying assumption that this ratio is to a certain degree model-independent,
based on the expectation that DM should probe the inner structure of nucleons in a similar way as neutrinos do
when only neutral current interactions are involved. Physically, indeed, this closely resembles the situation both for 
contact interactions and $t$-channel mediators.
The model dependence thus dominantly comes from the structure of the term 
$\left.\mathrm{d}\sigma_{\mathrm{SI}} / \mathrm{d}\omega\right|_{\mathrm{el}}$, and we approximate
\be
\label{eq:I2}
I_{\chi,i}(T_\chi,\omega) \approx I_{\nu,i}(E_\nu,\omega)\equiv\frac{\left.\mathrm{d}\sigma^i_{\nu N} /\mathrm{d}\omega \right|_{\mathrm{inel}}}
	{\mathrm{d}{\sigma}^i_{\nu,\mathrm{SI}} /\mathrm{d}\omega \big|_{\mathrm{el}}}\,.
\ee
Here, the inelastic neutrino-nucleus cross section 
$\left.\mathrm{d}\sigma_{\nu N}^i/\mathrm{d}\omega\right|_{\mathrm{inel}}(E_\nu,\omega)$ 
can be obtained using the \texttt{GiBUU} code, as described in section \ref{sec_gibuu}, and we evaluate it 
at the incoming DM kinetic energy, $E_\nu = T_\chi$.
 On the other hand, a possible estimate for the denominator -- the elastic neutral current neutrino-nucleon cross 
section without the form factor -- is the average of the proton and neutron cross sections in the $\omega \rightarrow 0$ 
limit~\cite{Formaggio:2012cpf}:
\begin{equation}
\label{eq:nuel_simp}
\left.\frac{\mathrm{d}\sigma^i_{\nu,\mathrm{SI}}}{\mathrm{d}\omega} \right|_{\mathrm{el}} 
= \frac{1}{2} \sum_{j=n,p} \frac{m_j G_F^2}{4\pi}\left[(g_A\tau_3^j - \Delta_S)^2 + (\tau_3^j-2(1+\tau_3^j)\sin^2\theta_W)^2\right].
\end{equation}
Here $\tau_3^p=1$ and $\tau_3^n=-1$, $\theta_W$ is the weak mixing angle and $G_F$ is the Fermi constant.
The axial vector and strange quark contributions are encoded in the parameters 
$\Delta_S\approx -0.15$ (see, e.g., Ref.~\cite{Alberico:1997vh} for a discussion) and  
$g_A= 1.267$~\cite{ParticleDataGroup:2008zun}, respectively. Numerically the square bracket evaluates
to a factor of $\sim\!2.24\,(2.01)$ for neutrons (protons). 
Let us stress, however, that this formula is valid only for energies 
relevant for inelastic scattering, $0.1\,\mathrm{GeV}\lesssim E_\nu\lesssim10$\,GeV. 
At much smaller energies, only the valence quarks contribute to the scattering, and we would instead have
\begin{equation}
\label{eq:nuel_simpNR}
\left.\frac{\mathrm{d}\sigma^i_{\nu,\mathrm{SI}}}{\mathrm{d}\omega} \right|_{\mathrm{el}} 
= \frac{m_n G_F^2}{4\pi}
\end{equation}
for neutrons, while the scattering on protons is strongly suppressed by a factor of 
$Q_W^2=(1-4\sin^2\theta_W)^2\approx0.012$.

It is worth noting that in principle, we could improve the assumption made in Eq.~(\ref{eq:I2})
for the quasi-elastic process, because there is a 
well-controlled understanding of the analytic QE cross section via the Llewellyn-Smith formalism (see section~V 
of~Ref.~\cite{Formaggio:2012cpf}). For clarity, we choose to take a consistent prescription across all inelastic processes, 
and we have checked that including the full QE cross section would only introduce an 
additional $\mathcal{O}(1)$ factor in the DM QE cross section.
For the numerical implementation in \ds, we pre-tabulate $I_{\nu,i}$ from $T_\chi=0.01$\,GeV up to energies of $T_\chi=10$\,GeV,
with $200$ ($101$) equally log-spaced bins in $T_\chi$ ($\omega$)
and a normalization as given by Eq.~(\ref{eq:nuel_simp}),  and then interpolate between 
these values.\footnote{%
For significantly higher energies,  \texttt{GiBUU} is no longer numerically stable. Furthermore, 
the underlying equations that describe the interaction processes begin to fall outside their ranges of validity
as the $Z$ boson mass starts to get resolved.
At higher energies, where anyway only the DIS contribution is non-negligible,  a reasonable estimate 
can still be obtained by a simple extrapolation 
$I_{\nu,i}(T_\chi,\omega)\to I_{\nu,i}(T_\chi^{\rm ref},\omega^{\rm ref})$, 
with $\omega^{\rm ref} =\omega\,(T_\chi^{\rm ref}/T_\chi)^{0.25}$,  beyond some reference energy
$T_\chi^{\rm ref}\approx10$\,GeV\@. By running  \texttt{GiBUU} up to $E_\nu\sim30$\,GeV, we checked that
this prescription traces the peak location (in $\omega$) of the DIS contribution very well, 
independently of the exact choice of $T_\chi^{\rm ref}$. We also confirmed that 
the peak value of $I$ becomes roughly constant for such large energies. 
On the other hand, higher-order inelastic processes are expected to become increasingly important 
at very large energies, not covered in \texttt{GiBUU}.
We therefore only add the above extrapolation as an {\it option} in \ds, and instead completely cut the incoming CRDM
flux at $10$\,GeV in the default implementation. As a result, our bounds on the interaction strength
may be overly conservative for small DM masses $m_\chi\lesssim0.1$\,GeV.
}

We also must choose the reference values for the transferred momentum $Q^2_{i, \mathrm{ref}}$, which allows us to account 
for e.g.~mediators that may be much lighter than the electroweak scale. 
Importantly, each process (quasi-elastic, 
$\Delta$-resonance,...) is expected to have a different characteristic $Q^2$-$\omega$ dependence that takes into account the 
relevant binding energies and kinematic scaling. For example, in the case of elastic scattering, the 
relation $Q^2 = 2 m_N \omega$ holds, whilst for quasi-elastic processes, the relevant scattering component is a nucleon such 
that the cross section is peaked around $Q^2 \sim 2 \,{\overline{ m}}\,\omega$, where ${\overline m} \equiv (m_n + m_p)/2$. 
The resonance of a particle with mass $m_\mathrm{res}$ can be accounted for by noting that part of the transferred
kinetic energy is used to excite the resonance, such that the cross section peaks around 
$Q^2 \sim 2\, {\overline m}\, (\omega - (m_\mathrm{res} - \overline{m}))$. 
We have confirmed these expectations numerically by comparing directly to the 
doubly-differential cross section extracted from \texttt{GiBUU}.
From this numerical comparison we further extract that $Q^2 \sim 0.6\,\overline{m}\,(\omega \!-\! \omega_{\rm DIS})$,
with $\omega_{\rm DIS}=1.0$\,GeV, constitutes a very
good fit to the peak location of the DIS cross section.
In summary, we take the following reference values across the four inelastic processes:
\begin{align}
	Q^2_{\mathrm{QE}, \mathrm{ref}} = 2\, {\overline m} \omega \, \, , \,\,\,\, &Q^2_{\Delta, \mathrm{ref}} = 2\, {\overline m}\, (\omega - \Delta m_\Delta)  \nonumber \\
	Q^2_{\mathrm{res}, \mathrm{ref}} = 2\, {\overline m}\, (\omega - \Delta m_{\mathrm{res}}) \, \, , \,\,\,\, &Q^2_{\mathrm{DIS}, \mathrm{ref}} = 0.6\, {\overline m}\, (\omega - \omega_{\rm DIS})\,.
\end{align}
Here, $\Delta m_\Delta = 0.29 \, \mathrm{GeV}$ is the mass difference between the $\Delta$ baryon and an average nucleon, 
and $\Delta m_\mathrm{res} = 0.40 \, \mathrm{GeV}$ is an estimate for the corresponding average mass difference of the higher 
resonances (we checked that our final limits are insensitive to the exact value taken here).

\begin{figure}[t]
	\begin{center}
	\includegraphics[width=0.9\textwidth]{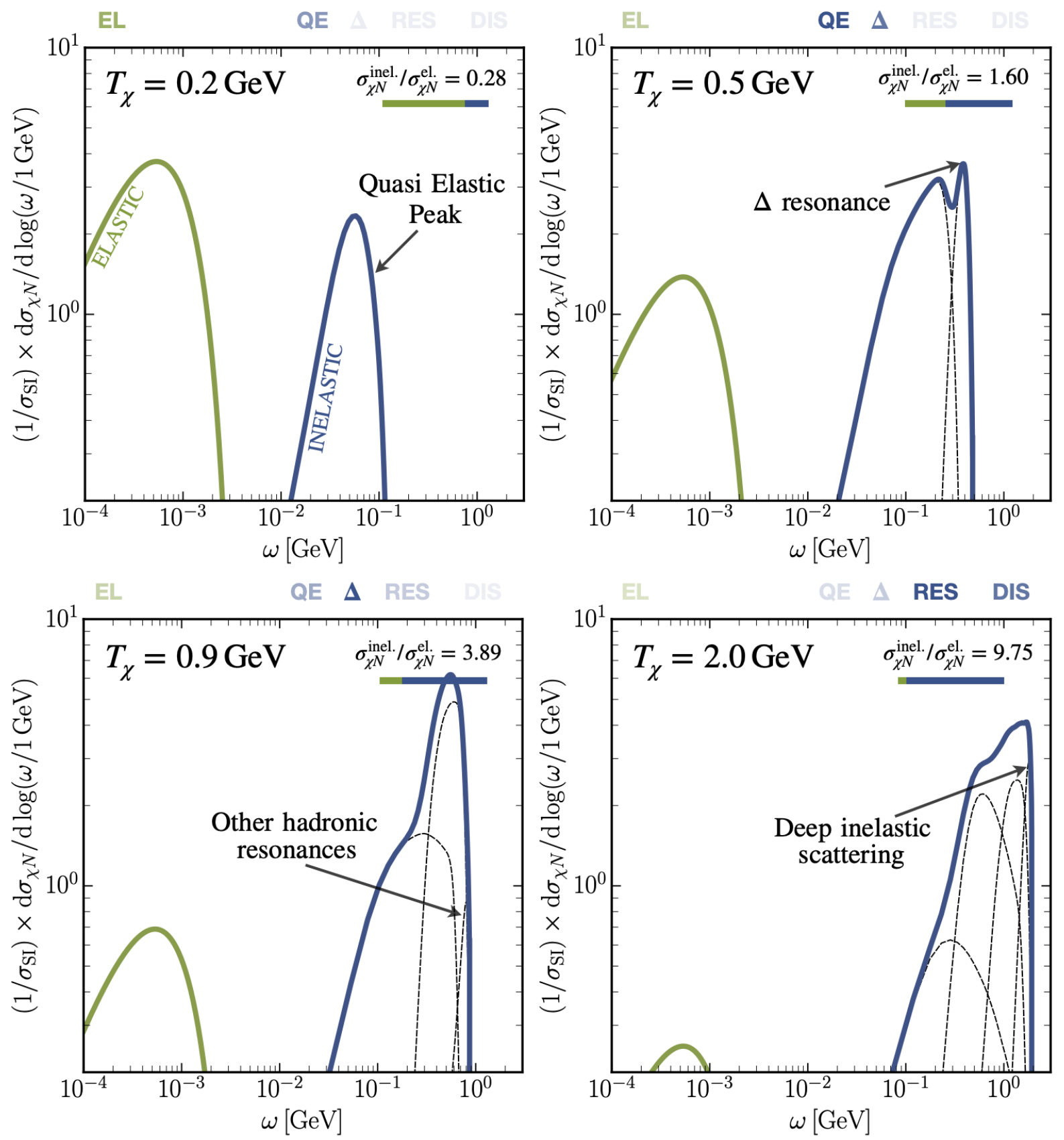}
	\caption{Comparison between the elastic (green, lower energies) and inelastic (blue, higher energies) contributions to the 
	DM-nucleus differential cross section $\mathrm{d}\sigma_{\chi N}/\mathrm{d}\omega$, where $\omega$ is the 
	DM energy loss. This figure shows these contributions 	for a constant isospin-conserving DM-nucleus cross section, with 
	$m_\chi = 1\, \mathrm{GeV}$ and $N = {^{16}}\mathrm{O}$. The small colorbar on 
	the inset of the plots, along with the stated numerical ratio, indicates the balance between elastic and inelastic
	scattering in terms of the contribution to the integrated cross section 
	$\sigma_{\chi N}^\mathrm{tot}$.
	}
	\label{fig:mchi_1GeV}
	\end{center}
\end{figure}

To illustrate this procedure concretely, we consider the simple case of a contact interaction where, cf.~Eq.~(\ref{eq:siconst}), 
$\left.\mathrm{d}\sigma_{{\rm SI}}/\mathrm{d}\omega\right|_{\mathrm{el.}} = \sigma_{\mathrm{SI}} / \omega^\mathrm{max}$ and 
$\omega^\mathrm{max} = 2\, {\overline m} (T_\chi^2 + 2 \chi T_\chi) / (({\overline m} + m_\chi)^2 + 2 {\overline m} T_\chi)$. The results for the 
rescaled inelastic cross section (blue) are shown in Fig.~\ref{fig:mchi_1GeV} for a DM mass $m_\chi = 1\,\mathrm{GeV}$ incident 
on a $^{16}\mathrm{O}$ nucleus. In this figure, we also compare to the coherent elastic contribution (green) and highlight the 
balance between the relative contributions to the total (integrated) cross section $\sigma^\mathrm{tot}_{\chi N}$. In particular, we 
see that above kinetic energies $T_\chi \gtrsim 0.2\,\mathrm{GeV}$, the inelastic contribution dominates, clearly 
motivating the necessity of its  inclusion. This is consistent with the picture previously encountered in Fig.~\ref{fig:attenuation_ff}, 
where we could see the impact of inelastic scattering on the energy loss.  More concretely, 
the result lies in some intermediate regime between the $G(Q^2) = 1$ and $G(Q^2) \neq 1$ cases, the former/latter leading 
to conservative/overly optimistic limits respectively. 
In the next section we will derive the relevant CRDM limits in the 
$\sigma_{\mathrm{SI}}-m_\chi$ plane for a number of models to make this point quantitatively.

Let us conclude this section by briefly returning to the implicit assumption of isospin-conserving DM interactions that
we made above, with 
$\sigma_{\mathrm{SI}}=\sigma^p_{\mathrm{SI}}=\sigma^n_{\mathrm{SI}}$.
Interestingly, neutral-current induced
inelastic scatterings between neutrinos and nucleons hardly distinguish between protons and 
neutrons~\cite{Formaggio:2012cpf}, such that the factor $I_{\chi,i}\approx I_{\nu,i}$  indeed becomes, by construction,
largely independent of the nucleon nature. Naively, one would thus conclude that
isospin-violating DM couplings can easily be incorporated in our treatment of inelastic scattering by replacing
$\sigma_{\mathrm{SI}} \to (1 / A) \times (Z \sigma^p_{\mathrm{SI}} + (A - Z) \sigma^n_{\mathrm{SI}})$  in Eq.~(\ref{eq:rescaling}).
When doing so, however, it is important to keep in mind that the nucleon cross sections should be evaluated at 
energies that are relevant for inelastic scattering, not  in the highly non-relativistic limit.
At these high energies, isospin symmetry is typically largely restored because the nucleon couplings are no longer 
exclusively determined by the valence quarks, and instead receive corrections from a large number of sea quarks 
(and, in principle, gluons).
As pointed out above, the example of neutrino scattering illustrates this effect very clearly: even though isospin is almost
maximally violated at low energies, the effective neutrino couplings to neutrons and protons agree within
$\sim5$\,\% at energies around 0.1\,GeV, cf.~Eqs.~(\ref{eq:nuel_simp}) and (\ref{eq:nuel_simpNR}). In practice, however,
a possible complication often arises in that the nucleon couplings $g_n$ and $g_p$ are only provided in the highly
non-relativistic limit. 
In that case, an educated guess for $\sigma_{\rm SI}$ in the second term of Eq.~(\ref{eq:rescaling})
is to anyway take the leading order (Born) expression -- but to adopt (effective)
values for {\it both} nucleon couplings that correspond to
the maximum of $\left|g_p\right|$ and $\left|g_n\right|$ in the non-relativistic limit. 
This induces a model-dependent uncertainty 
in the normalization of the inelastic contribution that can in principle only be avoided by fully implementing the concrete
interaction model in a code like \texttt{GiBUU}. On the other hand, the neutrino example illustrates that this error should
generally not be expected to be larger than a factor of $\sim$\,2, implying that for most applications such a 
more sophisticated treatment is not warranted.

\section{Contact interactions and beyond}
\label{sec:m2}

In sections \ref{sec:form_factors} and \ref{sec:inel} we have discussed in detail the 
$Q^2$-dependence that arises due to both form factor suppression and inelastic
scattering, as well as the impact this has on the production and attenuation of the CRDM flux.
This does not yet take into account, however, the possible angular and energy dependence of the
elastic scattering cross section itself. In fact, for \mbox{(sub-)GeV} DM, a significant dependence of this type is 
actually expected in view of null searches for new light particles at colliders. For example, it has been 
demonstrated in a recent global analysis~\cite{GAMBIT:2021rlp} that it is impossible to satisfy all relevant 
constraints simultaneously (even well above GeV DM masses) 
and at the same time maintain the validity of an effective field theory description at LHC energies.

Of course, this necessarily introduces a model-dependent element to the discussion, and in this section, the
aim will be to analyse the most generic situations that can appear when considering models beyond simple contact 
interactions.  Concretely, in section~\ref{sec:scalar} we will study the case of a light scalar mediator, a light vector
mediator in section~\ref{sec:vector}, and the scenario where DM particles have a finite extent
in section~\ref{sec:puffy}. In all these cases, we will re-interpret the published Xenon-1T limits and assess 
whether there is a remaining unconstrained window of large scattering cross sections for GeV-scale DM. 
Just before this, however, in section~\ref{sec:const} we will briefly revisit the (physically less motivated) case of 
a constant cross section, which can be viewed as the highly non-relativistic limit of a contact interaction. 
This will allow us to illustrate how the resulting CRDM constraints compare with 
established bounds from both surface and astrophysical experiments, as well as provide a more direct comparison with the 
existing literature.

\subsection{Constant cross section}
\label{sec:const}

\begin{figure}[t]
\begin{center}
\includegraphics[width=0.99\textwidth]{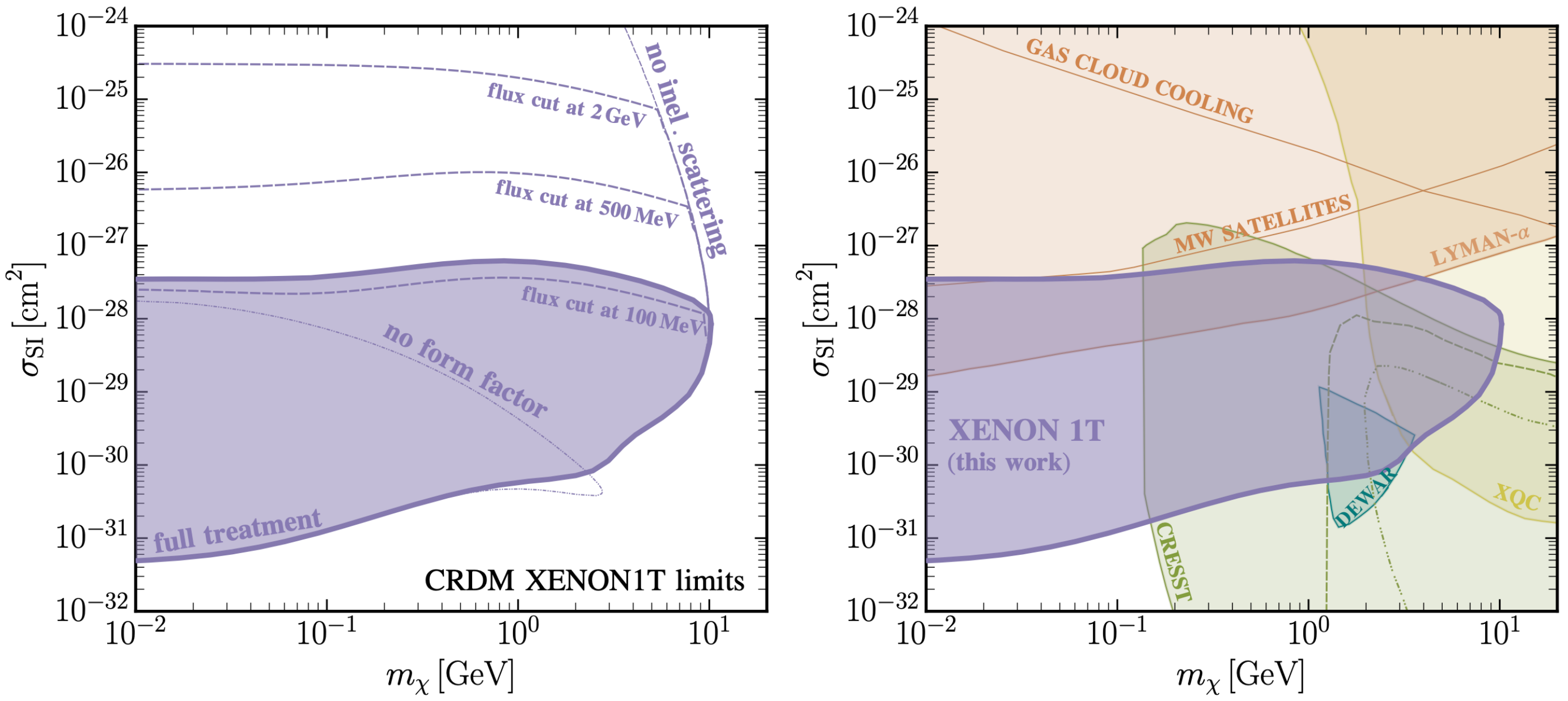}
\caption{%
{\it Left panel.} Limits on a constant spin-independent DM-nucleon scattering
cross section as a function of the DM mass, 
based on a re-interpretation of Xenon-1T limits on non-relativistic DM~\cite{XENON:2018voc}
for the CRDM component studied in this work (solid lines).
Dash-dotted lines show the excluded region that results when assuming a constant cross
section in the attenuation part (as in Ref.~\cite{Bringmann:2018cvk}).
Dashed lines show the effects of adding form factors in the attenuation part, but no 
inelastic scattering, resulting in limits similar to those derived in Ref.~\cite{Xia:2021vbz}.
For the latter case, for comparison, we also show the effect of artificially cutting the incoming CRDM flux 
at the indicated energies.\\ 
{\it Right panel.} Updated CRDM limits (coinciding with the solid lines from the left panel) in comparison 
to limits from the Lyman-$\alpha$ forest~\cite{Rogers:2021byl}, the Milky Way satellite 
population~\cite{Maamari:2020aqz}, gas clouds in the Galactic Centre 
region~\cite{Bhoonah:2018gjb}, the XQC experiment~\cite{McCammon:2002gb,Mahdawi:2018euy}, and 
a recently analysed storage dewar experiment~\cite{Neufeld:2019xes,Xu:2021lmg}.
We also show upper limits on the cross section as published by the CRESST 
collaboration~\cite{CRESST:2017ues} (solid green lines), based on a surface run of their 
experiment, along with the maximal cross section where 
attenuation does not prevent DM from leaving a signal in the detector~\cite{Emken:2018run}.
Alternative limits are indicated by green dashed~\cite{Mahdawi:2018euy} 
and dash-dotted lines~\cite{Xu:2020qjk}, based on the assumption of a thermalization efficiency 
of $\epsilon_{\rm th}=2$\,\% and $\epsilon_{\rm th}=1$\,\%, respectively, which is 
significantly worse than the one adopted in the CRESST analysis.
}
\label{fig:constraints_constant}
\end{center}
\end{figure}

For the discussion of a constant cross section, we will again consider the case of spin-independent scattering with isospin 
conserving nucleon couplings, cf.~Eq.~(\ref{eq:siconst}). In the left panel of Fig.~\ref{fig:constraints_constant}, we 
show our improved constraints from a re-interpretation of the Xenon-1T limits in this case.
Broadly, these updated and refined CRDM limits cover the mass range up to 
$m_\chi \lesssim 10 \,\mathrm{GeV}$ for cross sections 
$10^{-31}\,\mathrm{cm}^2 \lesssim \sigma_{\mathrm{SI}} \lesssim 2 \times 10^{-28}\,\mathrm{cm}^2$.

For comparison, we also indicate (with dash-dotted lines) the limits that result when neglecting both form-factor
dependence of the cross section and inelastic scatterings in the attenuation part. As expected, this leads to a shape of the 
excluded region very similar to that originally derived in Ref.~\cite{Bringmann:2018cvk}, where the same simplifying
assumptions were made. As a result of our improved treatment of CR fluxes and form factors, 
however, the limits indicated with dash-dotted lines are overall  slightly more stringent than what is reported in that analysis.
We find that for very light DM, with $m_\chi\lesssim10$\,MeV, this simplistic treatment actually leads to rather
realistic limits, the reason being that for highly relativistic particles the typical momentum transfer is always so large
that efficient inelastic scattering becomes relevant. For heavier DM masses, on the other hand, this treatment clearly 
overestimates the stopping power because it neglects the form factor suppression  relevant for semi-relativistic DM
scattering on nuclei.

Dashed lines furthermore show the effect of adding the form factor suppression during the attenuation in the soil, as done in 
Ref.~\cite{Xia:2021vbz}, but still not including inelastic scattering. Clearly, this vastly underestimates the actual
attenuation taking place and therefore appears to exclude very large cross sections.\footnote{%
Compared to Ref.~\cite{Xia:2021vbz}, we also find that the excluded region extends to somewhat larger DM masses,
mostly as a result of our updated treatment of elastic form factors.
On the other hand, we recall that our attenuation prescription is based on the analytical energy loss treatment outlined in 
section~\ref{sec:crdm}, rather than a full Monte Carlo simulation. This likely overestimates the maximally excluded DM mass,
but only by less than a factor of 2~\cite{Xia:2021vbz}.
} 
In order to gain a better intuitive understanding for the shape and strength of our final limits, finally, we also indicate the 
effect of neglecting inelastic scattering and instead artificially cutting the CRDM flux (prior to entering the soil) above 
some given energy.  The resulting upper limit on the cross section that can be probed in this fiducial setup strongly suggests that 
inelastic scattering events very efficiently stop the incident CRDM flux in the overburden as soon as they become
relevant compared to elastic scattering events. From Fig.~\ref{fig:constraints_constant}, and well in accordance with the
expectations from section \ref{sec:inel}, this happens at CRDM energies $T_\chi\gtrsim0.2$\,GeV.

In the right panel of Fig.~\ref{fig:constraints_constant} we show our improved constraints 
from a re-interpretation of the Xenon-1T limits in comparison with complementary limits from direct
probes of the DM-nucleon scattering cross section.
At small DM masses the dominant
constraint results from analysing the distribution of large-scale structures as traced by the Lyman-$\alpha$ 
forest. This is based on the fact that protons scattering too strongly off DM 
would accelerate the latter and thereby suppress the matter power spectrum at sub-Mpc scales. 
Such limits have recently been significantly tightened~\cite{Rogers:2021byl},  utilizing
state-of-the-art cosmological hydrodynamical simulations of the intergalactic medium at redshifts 
$2\lesssim z\lesssim6$. Similar bounds from the CMB (not shown here) are generally weaker by up to three 
orders of magnitude~\cite{Rogers:2021byl, Planck:2015bpv, Xu:2018efh}, while
the Milky Way satellite population~\cite{Maamari:2020aqz} -- as inferred from the Dark Energy Survey and 
PanSTARRS-1~\cite{DES:2019vzn} -- places bounds that are roughly one order of magnitude weaker.
Beyond cosmological bounds, cold gas clouds near the Galactic Center provide an interesting complementary testbed, 
in particular at high DM masses, where halo DM particles scattering too efficiently on the much {\it colder} baryon population 
would heat up the latter~\cite{Bhoonah:2018wmw}. Here we show updated 
constraints~\cite{Bhoonah:2018gjb} based on 
the cloud G357.8-4.7-55, noting that these constraints might be improved by more than one order
of magnitude if G1.4-1.8+87 is indeed as cold as $T\leq22$\,K (as reported in 
Refs.~\cite{McClure-Griffiths:2013awa,DiTeodoro:2018ybg} but disputed in Ref.~\cite{Farrar:2019qrv}).
We also display the limits~\cite{Mahdawi:2018euy} that result from the ten minutes' flight of the X-ray Calorimetry 
Rocket (XQC)~\cite{McCammon:2002gb}, based on the observation that ambient DM particles scattering
off the silicon nuclei in the quantum calorimeter would deposit (part of) their energy in the 
process~\cite{Wandelt:2000ad,Zaharijas:2004jv,Erickcek:2007jv}. 
In deriving these XQC limits, one must take into account that the recoil energy of a silicon nucleus 
potentially thermalizes much less efficiently in the calorimeter than the $e^\pm$ pairs produced from 
an incoming X-ray photon, such that a nuclear recoil energy $T_N$ will leave a signal equivalent  
to a photon with a reduced `thermal' recoil energy $T_T=\epsilon_{\rm th} T_N$. Concretely, the limits shown in the 
plot are based on the very conservative assumption of a thermalization efficiency factor of $\epsilon_{\rm th} = 0.02$.\footnote{%
When the scattering is mediated by a Yukawa-like interaction, a perturbative description of the scattering
process may no longer be adequate. In that case the constraints shown here, in particular for XQC, receive 
corrections due to non-perturbative effects leading to resonances or anti-resonances in the scattering cross 
section~\cite{Xu:2020qjk}. Here, we will not consider this possibility further, noting that a variation of the relatively 
uncertain value of $\epsilon_{\rm th}$ anyway has a larger impact on the XQC limits~\cite{Mahdawi:2018euy}.
}

Furthermore, in order to directly probe sub-GeV DM with very large cross sections, the CRESST collaboration has performed
a dedicated surface run of their experiment~\cite{CRESST:2017ues}, deliberately avoiding the shielding 
of the Gran Sasso rock used in the standard run~\cite{CRESST:2015txj}. The result of this search is the exclusion region 
indicated by the solid green line in Fig.~\ref{fig:constraints_constant}. Here, upper bounds on the cross section correspond to 
the published limits, obtained under the assumption that any attenuation in the overburden can be neglected. 
Modelling the effect of attenuation with detailed numerical simulations also results in  
the exclusion region limited from above~\cite{Emken:2018run}, coming from the fact that one must have a sufficiently large flux of 
DM particles at the detector location.
In a series of papers, Farrar~\textit{et al.}~have claimed  
that the CRESST thermalization efficiency adopted in the official analysis is too optimistic~\cite{Mahdawi:2018euy,Wadekar:2019mpc,Xu:2020qjk,Xu:2021lmg}, challenging the general ability of the experiment to probe sub-GeV DM.
We indicate the resulting alternatives to the published CRESST limits in the same figure, albeit noting that the underlying assumption 
of an efficiency as low as $\epsilon_{\rm th}\sim1$\,\% is not supported by data or simulations.
For example, no indication for such a dramatic loss of efficiency at low energies is observed for neutrons from an AmBe 
neutron calibration source~\cite{florian}.

To summarise, Fig.~\ref{fig:constraints_constant} illustrates the fact that the existence of the CRDM component provides an 
important probe of strongly interacting light DM. In particular, below $m_\chi\lesssim100$\,MeV, it restricts parameter space that 
is otherwise either unconstrained or only testable with 
cosmological probes (which -- at least to some degree -- are 
subject to modelling caveats regarding the Lyman-$\alpha$ forest and the non-linear evolution of density perturbations 
at small scales; see, e.g., Refs.~\cite{Hui:2016ltb,Irsic:2017ixq}). The CRDM component also leads to
highly relevant complementary constraints up to DM masses of a few GeV, especially when noting that  
these constraints are independent of the thermalization efficiency discussion above.

\subsection{Scalar mediators}
\label{sec:scalar}

As our first example beyond a constant scattering cross section we consider the case where 
a new light scalar particle $\phi$ mediates the interaction between DM and nucleons.  
We thus consider the interaction Lagrangian 
\be
\mathcal{L}_{\rm int}= - g_\chi \phi\overline\chi\chi - g_p \phi \overline p p- g_n \phi \overline n n\,,
\ee
and assume, for simplicity, isospin conservation ($g_p = g_n $).
At the level of the effective nuclear interaction Lagrangian, the dominant interaction terms with scalar ($N_0$) and fermionic ($N_{1/2}$) 
{\it nuclei} are thus given by\footnote{%
While the dominant cosmic-ray nuclei are either scalar or spin $1/2$ particles, some heavier nuclei in the overburden
have higher spins. For simplicity we treat those nuclei as scalars when determining their contribution to the energy loss,
as described by Eq.~(\ref{eq:eloss}), noting that this induces a neglible error in the estimated elastic scattering cross section,
of the order of $Q^2/m_N^2\ll1$. Moreover, nuclei with higher spins make up only about 2\% of the total mass in the overburden.
}
\be
\label{eq:leff_scalar}
\mathcal{L}_{\rm int}= -g_N\left( 2m_N N_0N_0+\overline N_{1/2}N_{1/2}\right)\,.
\ee
Here, the dimensionful coupling to scalar nuclei has been normalized such that both terms in the above expression result
in the same scattering cross section in the highly non-relativistic limit. In addition, the coupling to individual nucleons is coherently 
enhanced across the nucleus, resulting in an effective coupling to both scalar and fermionic nuclei given by
\be
\label{eq:gN_coh}
g_N^2 =A^2 \, g_p^2 \times G_N^2(Q^2)\,,
\ee
where $G_N$ is the same form-factor as in the case of a `constant' cross section.
For the resulting elastic scattering cross section for DM incident on nuclei at rest we find
\be
\label{diffsig_full_scalar}
\frac{d\sigma_{\chi N}}{d T_N}=\frac{\mathcal{C}^2\sigma_{\rm SI}^\mathrm{NR}}{T_N^\mathrm{max}}
\frac{m_\phi^4}{(Q^2+m_\phi^2)^2}
\frac{m_N^2\left(Q^2+4m_\chi^2 \right)}{4 s\,{\mu_{\chi N}^2}}
\times\left\{
\begin{array}{ll}
1& ~~\mathrm{for~scalar~}N\\
1+\frac{Q^2}{4m_N^2} &  ~~\mathrm{for~fermionic~}N
\end{array}
\right\}
\times G_N^2(Q^2)\,,
\ee
where $\mu_{\chi p}$ is the reduced mass of the DM/nucleon system and  
\be
\label{eq:sig0_scalar}
\sigma_{\rm SI}^\mathrm{NR} = \frac{g_\chi^2 g_p^2 \mu_{\chi p}^2}{\pi m_\phi^4}
\ee
is the spin-independent scattering cross section {\it per nucleon} in the ultra non-relativistic limit.
For reference, the kinematic quantities $T_N^\mathrm{max}$, $s$ and $Q^2$ are given by 
Eqs.~(\ref{eq:tmax}), (\ref{eq:sdef}) and (\ref{eq:q2}), respectively. For the production part of the process, 
where CR nuclei collide with DM at rest, one
simply has to exchange $T_N\leftrightarrow T_\chi$ and $m_\chi\leftrightarrow m_N$ in 
these expressions for kinematic variables -- but not in the rest of Eq.~(\ref{diffsig_full_scalar}) 
-- in order to obtain ${d\sigma_{\chi N}}/{d T_\chi}$.

\begin{figure}[t]
\begin{center}
\includegraphics[width=0.99\textwidth]{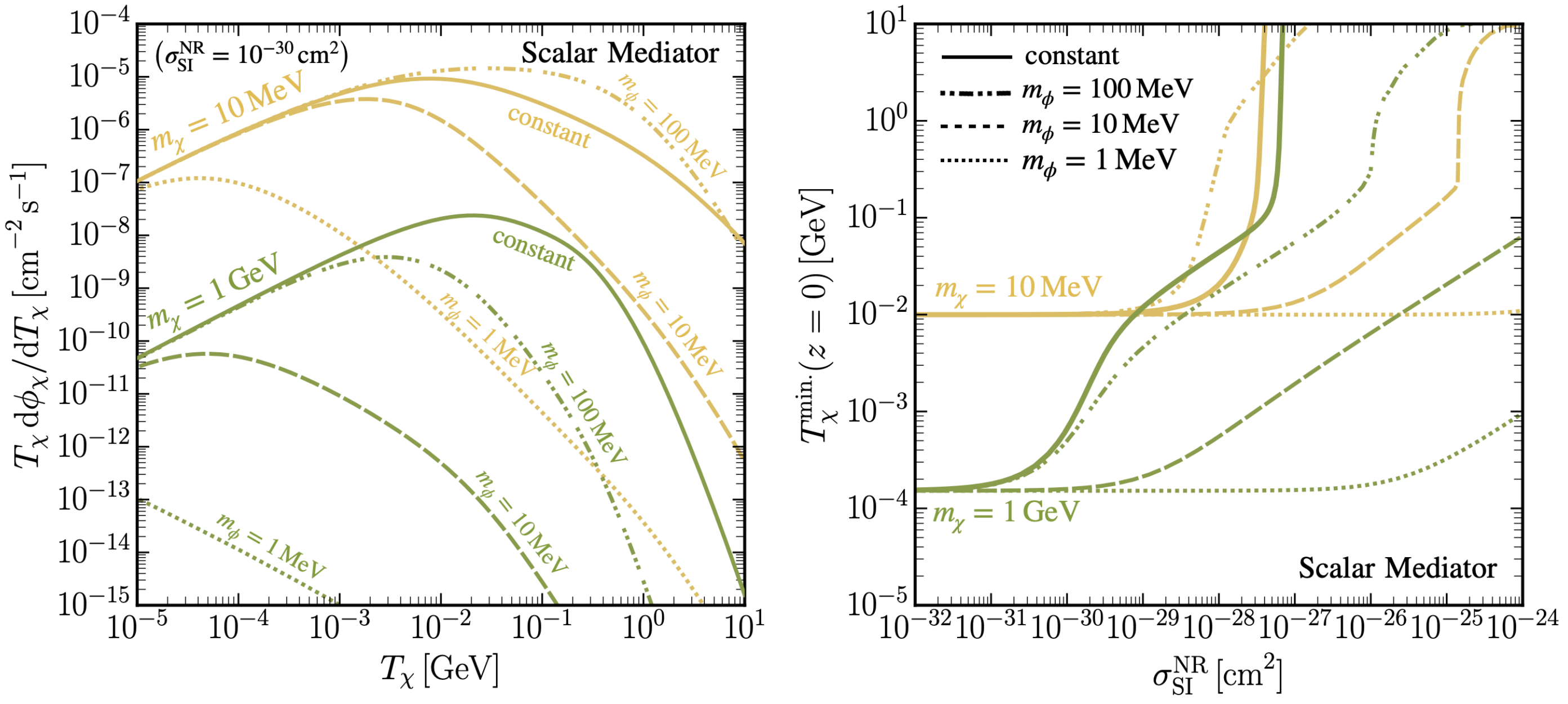}
\caption{{\it Left panel.} Solid lines show the CRDM flux before attenuation for a constant interaction
cross section, as in Fig.~\ref{fig:flux_ff}, for DM masses  $m_\chi=10$\,MeV and  $m_\chi=1$\,GeV. 
For comparison we also indicate the corresponding CRDM flux for a {\it scalar} mediator, cf.~Eq.~(\ref{diffsig_full_scalar}), 
with mass  $m_\phi=100$\,MeV (dash-dotted), $m_\phi=10$\,MeV 
(dashed) and $m_\phi=1$\,MeV (dotted), for a cross section (in the non-relativistic limit) of  
$\sigma_{\rm SI}^{\rm NR}=10^{-30}\,\mathrm{cm}^2$. 
{\it Right panel.} 
Minimal kinetic energy $T_\chi$ that a DM particle must have, prior to attenuation, in order 
to trigger a signal in the Xenon-1T experiment. Line styles and colors match those of the left
panel. In particular, solid lines show the case of a constant spin-independent scattering cross 
section and are identical to those displayed  in Fig.~\ref{fig:attenuation_ff}.
}
\label{fig:results_scalar}
\end{center}
\end{figure}

In the left panel of Fig.~\ref{fig:results_scalar} we show the resulting CRDM fluxes for this
model. For small kinetic energies these fluxes are, as expected, identical to those shown in Fig.~\ref{fig:flux_ff} 
for the case of a constant cross section. This is the regime where $Q^2=2m_\chi T_\chi$ is smaller
than the masses of both the mediator and CR nuclei, such that Eq.~(\ref{diffsig_full_scalar}) reduces to 
Eq.~(\ref{eq:siconst}). For $Q^2\gtrsim m_\phi^2$, on the other hand, the presence of a light mediator 
clearly suppresses the fluxes. Note that the matrix element also contains a factor of $(Q^2+4m_\chi^2)$, 
which additionally leads
to a flux {\it enhancement} for fully relativistic DM particles, $T_\chi\gtrsim 2 m_\chi$. 
In the figure, this latter effect is clearly visible for the case of $m_\chi=10$\,MeV and a heavy mediator.
In general, the appearance of such model-dependent features demonstrates the need to use the full matrix 
element for the relativistic cross section. This is in contrast to the non-relativistic case, where a model-independent rescaling 
of the cross section by a factor of $(1+Q^2/m_\phi^2)^{-2}$ is usually sufficient to model the effect of a light mediator
(see, e.g., Refs.~\cite{Chang:2009yt,Fornengo:2011sz,Kaplinghat:2013yxa}).

In the right panel of Fig.~\ref{fig:results_scalar}, we explore the minimal CRDM energy $T_\chi$ that
is needed to induce a detectable nuclear recoil. Compared to the situation of a constant scattering cross section (depicted by the 
solid lines for easy comparison), the attenuation is as expected rather strongly suppressed when light scalar mediators are 
present (with the exception of the
case with $m_\chi=10$\,MeV and $m_\phi=100$\,MeV, where the cross section is enhanced 
due to the $(Q^2+4m_\chi^2)$ factor in the squared matrix element). In order to understand the qualitative behaviour of 
$T_\chi^{\rm min} (z=0)$ better, we recall from the discussion of Fig.~\ref{fig:attenuation_ff}
that there are two generic scaling regimes for solutions of the energy loss equation. Firstly, for cross sections with no 
-- or only a mild --  dependence on the momentum transfer, $T_{\chi}^{\rm min}(z=0)$ grows exponentially 
with increasing $\sigma_{\rm SI}^{\rm NR}$. Secondly, in the presence of an effective cutoff in the cross section (like when form 
factors or light mediators are introduced), $T_{\chi}^{\rm min}(z=0) \propto \sqrt{\sigma_{\rm SI}^{\rm NR}}$ for large energies
$T_\chi$. These different regimes are clearly visible in the figure. 
For the green dot-dashed curve ($m_\chi=1$\,GeV, $m_\phi=100$\,MeV), for example,
one observes as expected an initial steep rise at the smallest DM energies -- until the form factor and mediator suppression 
of the cross section cause a scaling with $\sqrt{\sigma_{\rm SI}^{\rm NR}}$ for kinetic energies above a few MeV. 
At roughly $T_\chi\gtrsim0.1$\,GeV, inelastic scattering kicks in, leading again to an exponential suppression of the flux. 
For even higher energies, finally, the scattering cross section falls off so rapidly that the required initial DM energy once
again only grows as $\sqrt{\sigma_{\rm SI}^{\rm NR}}$.

Turning our attention to the resulting CRDM limits, it is worth stressing here that $\sigma_{\rm SI}^\mathrm{NR}$, as introduced in Eq.~(\ref{eq:sig0_scalar}),
is a somewhat artificial object that only describes the cross section for physical processes
restricted to $Q^2\lesssim m_\phi^2$.
In a direct detection experiment like Xenon-1T this is necessarily violated for $m_\phi\lesssim \sqrt{2m_N T_N^{\rm thr}}\sim35$\,MeV, 
given that  $T_N^{\rm thr}=4.9$\,keV is the  minimal recoil 
energy needed to generate a signal. A natural consequence of this is that making a straight-forward comparison 
to the $\sigma_{\rm SI}$ appearing in the `constant cross section' case discussed in 
section \ref{sec:const} is challenging. Instead, the best we can achieve in terms of a meaningful 
comparison is to define a {\it reference cross section}
\be
\label{eq:Qref}
 \tilde \sigma_{\rm Xe,SI}^p \equiv \sigma_{\rm SI}^\mathrm{NR}\times \frac{m_\phi^4}{(Q_{\rm Xe,ref}^2+m_\phi^2)^2}
\frac{Q^2_{\rm Xe,ref}+4m_\chi^2}{4m_\chi^2}\,,
\ee
where $Q_{\rm Xe,ref}\sim35$\,MeV. It follows from Eq.~(\ref{diffsig_full_scalar}) and Eq.~(\ref{eq:siconst}), 
and the fact that $s\approx (m_\chi+m_N)^2$ for the energies of interest here, that $ \tilde \sigma_{\rm Xe,SI}^p$ 
should be interpreted as the effective CRDM cross section per nucleon that is dominantly seen in the Xenon-1T 
analysis window.
It is thus this quantity, not the $\sigma_{\rm SI}^\mathrm{NR}$ from Eq.~(\ref{eq:sig0_scalar}), that should
be compared to the published Xenon-1T limits on the DM-nucleon cross section.

This also allows us to address the question of how the limits on the DM-nucleon coupling coming from the CRDM component
compare to the complementary constraints introduced in section \ref{sec:const} (cf.~the right
panel of Fig.~\ref{fig:constraints_constant}). In order to do so, one first needs to
realize that all of those limits are derived under the assumption of non-relativistic
DM and a constant cross section. In reality, however, they probe very different physical environments and typical 
momentum transfers. In order to allow for a direct comparison, therefore, they also need to be re-scaled to a common 
reference cross section.
Assuming that the DM energies in Eq.~(\ref{diffsig_full_scalar}) are non-relativistic, a reported limit on the DM-nucleon 
cross section $\sigma_{\rm SI}^p$ from an experiment probing typical momentum transfers of the order $Q^2_{\rm ref}$ would 
correspond to a cross section of
\be
\label{eq:rescale_scalar}
 \tilde \sigma_{\rm Xe,SI}^p = \sigma_{\rm SI}^p\times 
 \left(\frac{Q^2_{\rm ref}+m_\phi^2}{Q^2_{\rm Xe,ref}+m_\phi^2}\right)^2
\frac{Q^2_{\rm Xe,ref}+4m_\chi^2}{Q^2_{\rm ref}+4m_\chi^2}
\ee
in the Xenon-1T detector. As an example, consider the CRESST surface run~\cite{CRESST:2017ues}, where a threshold energy of $\sim20$\,eV
for the sapphire 
detector would imply $Q^2_{\rm ref}\sim (0.98\,\mathrm{MeV})^2/\epsilon_{\rm th}$.  Similarly, a thermal recoil energy of of $29$\,eV
in XQC corresponds to $Q^2_{\rm ref}\sim (8.7\,\mathrm{MeV})^2$ for the nuclear recoil on Si nuclei
(assuming $\epsilon_{\rm th} = 0.02$ as for the unscaled limits).
Turning to cosmological limits, a baryon velocity of $v_b^{\rm rms}\sim33$km/s at the times relevant for the emission
of Lyman-$\alpha$ photons~\cite{Silk:1967kq} implies typical momentum transfers from the Helium nuclei to DM of 
$Q_{\rm ref}^2\sim4 \mu_{\chi{\rm He}}^2\times10^{-8}$. This means that, for the range of DM and mediator masses considered 
here, the cross section at these times becomes roughly constant and we can approximate $Q_{\rm ref}^2\approx0$ 
in Eq.~(\ref{eq:rescale_scalar}). The same goes for the constraints stemming from the MW satellite abundance,
which are sensitive to even lower redshifts and thus smaller momentum transfers~\cite{Nadler:2019zrb,Maamari:2020aqz}.

\begin{figure}[t]
\begin{center}
\includegraphics[width=0.9\textwidth]{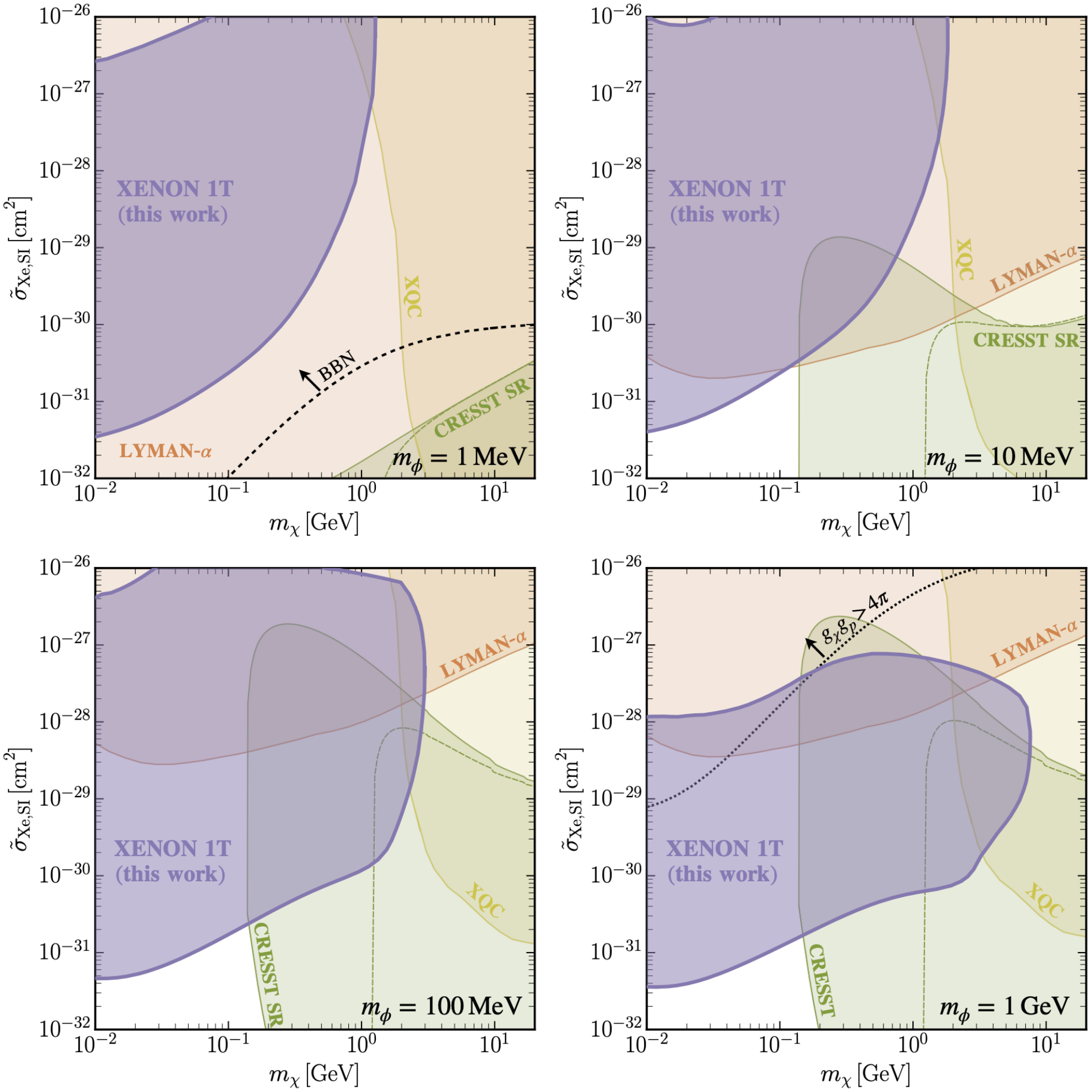}
\caption{%
Limits on the DM-nucleon scattering cross section evaluated at a reference momentum transfer of 
$Q_{\rm Xe,ref}=35$\,MeV, as a function of the DM mass $m_\chi$. From top left to bottom right,
the panels show the case of a {\it scalar mediator} with mass $m_\phi=1$\,MeV, 10\,MeV, 100\,MeV and 1\,GeV.
Solid purple lines show the updated CRDM limits studied in this work. We further
show limits from the Lyman-$\alpha$ forest~\cite{Rogers:2021byl}, the XQC 
experiment~\cite{McCammon:2002gb,Mahdawi:2018euy}, the CRESST surface 
run~\cite{CRESST:2017ues, Emken:2018run} and an alternative analysis of the CRESST 
limits~\cite{Mahdawi:2018euy}. All these limits are rescaled to match the situation of a light mediator,
as explained in the text. The parameter region above the dotted black line in the bottom right panel
requires non-perturbative couplings, while the area above the dotted line in the top left panel is excluded
by BBN.
}
\label{fig:limits_scalar}
\end{center}
\end{figure}

In Fig.~\ref{fig:limits_scalar} we show a subset of these correspondingly rescaled constraints\footnote{%
Upper bounds on the excluded cross section, due to attenuation effects, cannot simply be rescaled as
in Eq.~(\ref{eq:rescale_scalar}). 
For the sake of Fig.~\ref{fig:limits_scalar}, we instead adopt a rather simplistic 
approach~\cite{Davis:2017noy,Kouvaris:2014lpa,Emken:2017erx,Emken:2018run} to estimate these 
limits by requiring that the most energetic halo DM particles, with an assumed velocity $v_{\rm max}$,
can trigger nuclear recoils above the CRESST threshold of 19.7\,eV/$\epsilon_{\rm th}$ after attenuation in the 
Earth's atmosphere. For the average density and distribution of elements in the atmosphere, we follow Ref.~\cite{USatm}.
By treating $v_{\rm max}$ and the effective height of the atmosphere, $h_a$, 
as free parameters, we can then rather accurately fit the results of more detailed 
calculations~\cite{Emken:2018run,Mahdawi:2018euy} for the case of a constant cross section
-- with numerical values in reasonable agreement with the physical expectation in such a heuristic approach. 
Finally, we adopt those values of $v_{\rm max}$ and $h_a$ to derive the corresponding limits for the case of a scalar mediator,
as displayed in Fig.~\ref{fig:limits_scalar}.
}
 -- for mediator masses 
$m_\phi=1$\,MeV, 10\,MeV, 100\,MeV and 1\,GeV -- along with the full CRDM constraints derived here.
We also indicate, for comparison, with dotted black lines where non-perturbative couplings would be needed in this model to 
realize the stated cross section. This line is only visible for the case of $m_\phi=1$\,GeV,
which demonstrates that it is generically challenging to realize large cross sections without invoking light mediators.
The presence of an abundant species with a mass below a few MeV, furthermore, would affect how light elements are
produced during big bang nucleosynthesis (BBN). For a 1\,MeV particle with one degree of freedom, like $\phi$,  
this can be formulated as a constraint of $\tau>0.43$\,s~\cite{Depta:2020zbh} on the lifetime of such a particle.
Physically, this constraint derives from freeze-in production of $\phi$ via the inverse decay process. 
Since $\phi\to\gamma\gamma$ (apart from $\phi\to\bar \nu\nu$) is the only kinematically possible SM decay channel, 
the translation of this bound to a constraint on the SM coupling $g_p$ is somewhat model-dependent.
For concreteness we consider the Higgs portal model, where $\tau>1$\,s at $m_\phi=1$\,MeV corresponds to 
a squared mixing angle
$\sin^2\theta=(8.62\times10^2g_p)^2>3.8\times10^{-4}$~\cite{Krnjaic:2015mbs}. The area above the dashed
line in the top left panel of Fig.~\ref{fig:limits_scalar} requires either a {\it larger} value of $g_p$ than what is given by this
bound, or a non-perturbative coupling $g_\chi^2>4\pi$. This confirms the generic expectation that for very light particles 
BBN constraints are more stringent than those stemming from the CRDM 
component~\cite{Krnjaic:2019dzc,Bondarenko:2019vrb}.

Our results demonstrate that in the presence of light mediators the largest DM mass that can be constrained 
due to CR upscattering is reduced from about 10\,GeV, cf.~Fig.~\ref{fig:constraints_constant},
to just above 1\,GeV (for $m_\phi\sim1$\,MeV). This is a direct consequence of the suppressed CRDM production
rate discussed above. On the other hand, the reduction of the cross section also implies a smaller attenuation
effect, thus closing parameter space at larger cross sections. More importantly, complementary constraints
from cosmology and dedicated surface experiments become {\it more stringent} in the presence of light mediators, 
once they are translated to a common reference cross section. To put this in context, let us first recall that in the constant 
cross section case, Fig.~\ref{fig:constraints_constant} tells us that cross sections 
$\sigma_{\rm SI}\gtrsim 2\cdot10^{-31}\,{\rm cm}^2$  are safely excluded across the entire DM mass range 
(or $\sigma_{\rm SI}\gtrsim 6\cdot10^{-31}\,{\rm cm}^2$ when assuming that the thermalization efficiency of 
CRESST is indeed as low as 2\,\%). From Fig.~\ref{fig:limits_scalar} we infer that these limits can be somewhat 
weakened for sub-GeV DM, when considering light meditators in the mass range 
$10\,{\rm MeV}\lesssim m_\phi\lesssim 100\,{\rm MeV}$ (as we will see further down, the situation of a vector mediator
is not appreciably different from that of the scalar mediator shown here). Concretely, the upper bound on the 
cross section now becomes $\tilde \sigma_{\rm SI}\lesssim 3\cdot10^{-31}\,{\rm cm}^2$, independently of the DM
{\it and} mediator mass. For a 2\,\% thermalization efficiency of CRESST~\cite{Mahdawi:2018euy} and a narrow 
range of mediator masses, $10\,{\rm MeV}\lesssim m_\phi\ll100\,{\rm MeV}$,
a small window opens up above the maximal cross section that can be probed with CRESST. 
The reason is the gap between Lyman-$\alpha$ bounds and the weakened CRESST limits from Ref.~\cite{Mahdawi:2018euy}
that is visible in the figure, for $m_\phi\gtrsim10\,{\rm MeV}$, and which is closed by the CRDM limits only for mediator
masses of $m_\phi\gtrsim 30$\,MeV. Nominally, for $m_\chi\sim2$\,GeV and $m_\phi\sim 30$\,MeV, 
this would allow for cross sections as large as $\tilde \sigma_{\rm SI}\sim 4\cdot10^{-29}\,{\rm cm}^2$.
In either case, the conclusion remains that CRDM leads to highly complementary limits, and that this 
relativistic component of the DM flux is in fact crucial for excluding the possibility of very large DM-nucleon
interactions.

\subsection{Vector mediators}
\label{sec:vector}

We next consider the general case of a massive vector mediator $V$, with interactions given by
\be
\mathcal{L}=   V_\mu \left(g_\chi \overline{\chi}\gamma^\mu \chi + g_{p}\overline{p}\gamma^\mu p + g_{n}\overline{n}\gamma^\mu n\right)\,.
\ee
We will again assume  $g_n=g_p$  for simplicity, noting that smaller values of the ratio $g_n/g_p$ can lead to
significantly smaller cross sections (see, e.g., Refs.~\cite{Frandsen:2011cg,Kaplinghat:2013yxa}); in our context this would
mostly imply that the attenuation in the overburden becomes less relevant, leading to more stringent constraints.
In analogy to Eq.~(\ref{eq:leff_scalar}), this implies the following dominant interaction terms with scalar and fermionic nuclei, respectively:
\be
\label{eq:leff_vector}
\mathcal{L}_{\rm int}= -g_N V_\mu\left( i N_0^*{\mathop{\partial^\mu}^{\leftrightarrow}} N_0+\overline N_{1/2}\gamma^\mu N_{1/2}\right),
\ee
where the effective mediator coupling to nuclei, $g_N$, is again given by the coherent enhancement stated in Eq.~(\ref{eq:gN_coh}).
For the elastic scattering cross section on nuclei we find
\bea
\label{diffsig_full_vector}
\frac{d\sigma_{\chi N}}{d T_N}&=&\frac{\mathcal{C}^2 \sigma_{\rm SI}^\mathrm{NR}}{T_N^\mathrm{max}}
\frac{m_A^4}{(Q^2+m_A^2)^2}
\times G_N^2(Q^2)\\
&&
\times\frac{1}{4s \mu_{\chi N}^2}
\left\{
\begin{array}{ll}
m_\chi^2Q^2-Q^2s+(s-m_N^2-m_\chi^2)^2& ~~\mathrm{for~scalar~}N\\
\frac12 Q^4 -Q^2s+(s-m_N^2-m_\chi^2)^2&  ~~\mathrm{for~fermionic~}N
\end{array}
\right..\nonumber
\eea
Here, the cross section in the ultra-nonrelativistic limit,
\be
\label{eq:sig0_vector}
\sigma_{\rm SI}^\mathrm{NR} = \frac{g_\chi^2 g_p^2 \mu_{\chi p}^2}{\pi m_A^4}\,,
\ee
i.e.~for $Q^2\to0$ and $s\to(m_N+m_\chi)^2$, agrees exactly with the result obtained for the scalar case, as expected.
For large energies and momentum transfers, on the other hand, the behaviour
is different.

\begin{figure}[t]
\begin{center}
\includegraphics[width=0.99\textwidth]{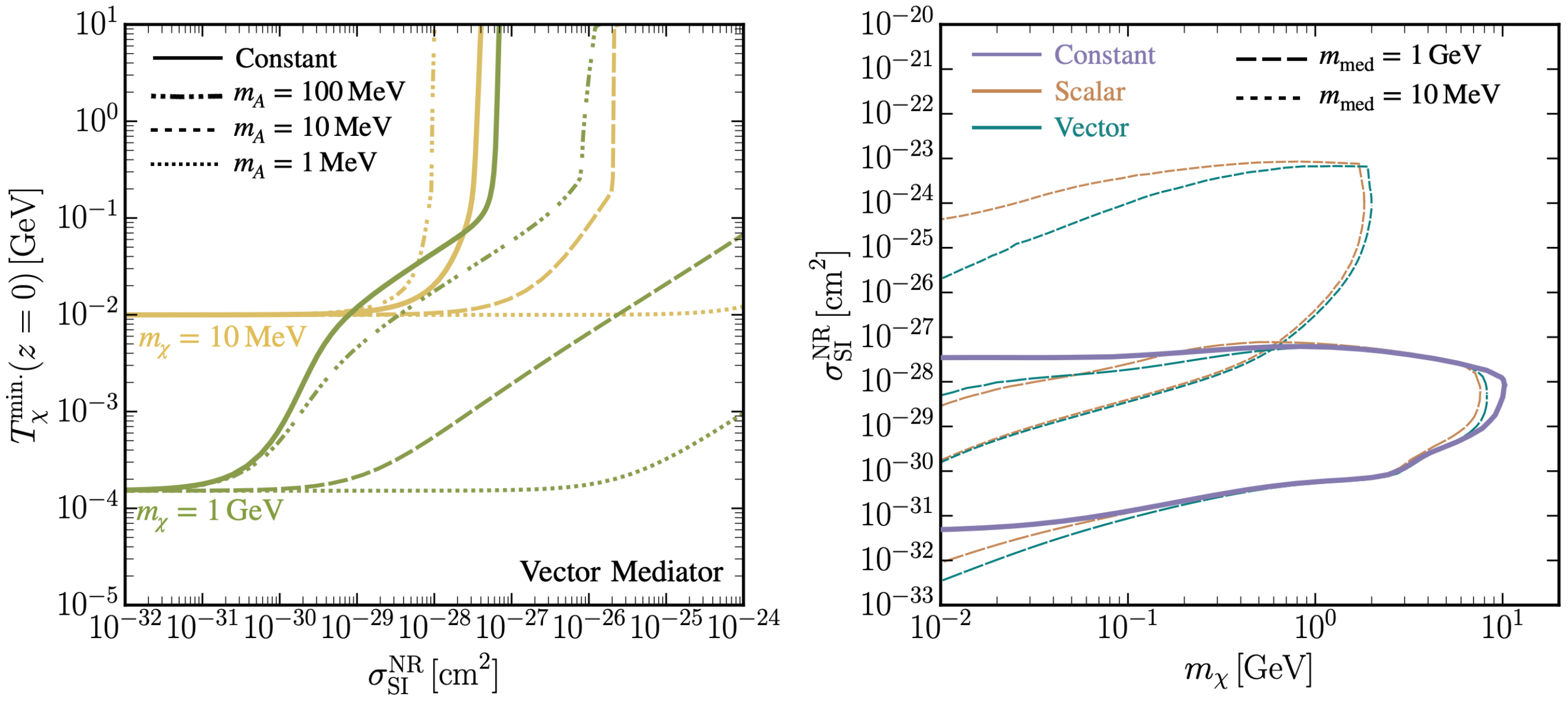}
\caption{%
{\it Left panel.} 
Minimal kinetic energy $T_\chi$ that a DM particle must have, prior to attenuation, in order 
to trigger a signal in the Xenon-1T experiment for DM nucleus interactions via a {\it vector mediator},
as a function of the spin-independent DM-nucleon scattering cross section in the highly non-relativistic limit, 
$\sigma_{\rm SI}^\mathrm{NR}$. Yellow (green) lines indicate a DM mass
$m_\chi=10$\,MeV ($m_\chi=1$\,GeV), and different line styles correspond to  mediator masses 
$m_A=1,10,100$\,MeV as indicated.
Solid lines show the case of a constant spin-independent scattering cross 
section and are identical to those displayed  in Fig.~\ref{fig:attenuation_ff}.
{\it Right panel.} 
Constraints on $\sigma_{\rm SI}^\mathrm{NR}$ as a function of the DM mass $m_\chi$.  Solid purple lines 
refer to the case of a constant cross section, as in Fig.~\ref{fig:constraints_constant}, while other line 
styles show the case where the interaction is mediated by a light scalar (red) or vector (green) particle 
with mass $m_{\rm med}=10$\,MeV and $1$\,GeV, respectively.}
\label{fig:limits_vector}
\end{center}
\end{figure}

The resulting CRDM fluxes are nonetheless so similar to the scalar case shown in the left panel of Fig.~\ref{fig:results_scalar} 
that we refrain from plotting them separately.  Differences do exist, however, for the stopping power in the overburden.
In the left panel of Fig.~\ref{fig:limits_vector} we therefore show the minimal initial kinetic energy needed by a CRDM
particle to induce detectable nuclear recoils in Xenon-1T. Compared to the scalar case, cf.~the right panel of 
Fig.~\ref{fig:results_scalar},  the attenuation is more efficient for highly relativistic DM particles due to the $s$-dependence 
of the terms in the second line of Eq.~(\ref{diffsig_full_vector}). As before, the effect of these model-dependent terms
from the scattering amplitude 
is most visible for highly relativistic particles, with small $m_\chi$, and large mediator masses, where 
the suppression due to the factor $(1+Q^2/m_A^2)^{-2}$ is less significant.

In the right panel of  Fig.~\ref{fig:limits_vector} we compare the  final exclusion regions for the 
situations considered so far, i.e.~for a contact interaction, scalar mediators and vector mediators, respectively. 
For the sake of comparison in one single figure, 
we plot here the cross section in the ultra-nonrelativistic limit. For an interpretation of 
these limits in comparison to complementary constraints on DM-nucleon interactions 
we thus refer to the discussion of Fig.~\ref{fig:limits_scalar},
noting that the rescaling prescriptions for vector and scalar mediators are qualitatively the same.
The first thing to take away from Fig.~\ref{fig:limits_vector} is that, as expected, the exclusion regions 
for heavy mediators resemble those obtained for the constant cross section case. The figure
further demonstrates that the only significant difference between scalar and vector mediators appears  
at smaller mediator masses, where the former are somewhat less efficiently stopped in the overburden. 
It is worth noting, however, that this region of parameter space where the vector and scalar cases differ substantially
is nonetheless excluded by Lyman-$\alpha$ bounds. 
The general discussion and conclusions from the scalar mediator case 
explored in the previous subsection thus also applies to interactions mediated by vector particles.

\subsection{Finite-size dark matter}
\label{sec:puffy}

As a final generic example of a $Q^2$-suppressed cross section let us consider the situation 
where the DM particle itself has a finite size that is larger than its Compton wavelength. 
Various models of such composite DM have been extensively studied in the 
literature~\cite{Nussinov:1985xr,Chivukula:1989qb,Cline:2013zca,Krnjaic:2014xza,Wise:2014ola,Hardy:2015boa,Coskuner:2018are,Contino:2018crt}.
In fact, Ref.~\cite{Digman:2019wdm} even suggests that DM with masses above 1\,GeV \emph{cannot} 
be point-like for DM-nucleon cross section $\gtrsim 10^{-25}\,\text{cm}^2$.
The corresponding scattering cross section then takes the same form as in the point-like case,
multiplied by another factor $\left|G_\chi(Q^2)\right|^2$ that reflects the spatial extent of 
$\chi$~\cite{Feldstein:2009tr,Laha:2013gva,Chu:2018faw}.
Specifically, just as for nuclear form factors, we have 
\be
G_\chi(Q^2)=\int d^3 x\,e^{i\mathbf{q}\cdot\mathbf{x}}\rho_\chi(\mathbf{x})\,,
\ee
where $\rho_\chi(\mathbf{x})$ is the distribution of the effective charge density that the interaction couples to.
For simplicity we will  choose a dipole form factor of the form\footnote{%
The exact choice of the form factor does not significantly affect our results, as long as $G_\chi(Q^2)<G_\chi(0)=1$.
An interesting, qualitatively different situation occurs when $G_\chi(0)=0$, i.e.~for a form factor that {\it grows} 
with $Q^2$. This is, e.g., realized if the scattering is mediated by a dark $U(1)'$ under which $\chi$ is 
neutral~\cite{Feldstein:2009tr,Chu:2018faw}. We will not consider this class of models in this work.
}
\be \label{eq:dipole}
G_\chi(Q^2)=\left(1+\frac{r_\chi^2}{12}Q^2\right)^{-2}\,,
\ee
with $r_\chi$ being the r.m.s.~radius of the DM particle, $r_\chi^2=\int d^3 x\,\mathbf{x}^2 \rho_\chi(\mathbf{x})$.
We then multiply $G^2_\chi(Q^2)$ with Eq.~(\ref{eq:siconst}) in order to obtain 
${d\sigma_{\chi N}}/{d T_N}$, thus describing an effective scalar interaction with the usual coherent
enhancement inside the  nucleus -- but where each of the nucleons only `sees' some fraction of the entire DM 
particle.

\begin{figure}[t]
\begin{center}
\includegraphics[width=0.99\textwidth]{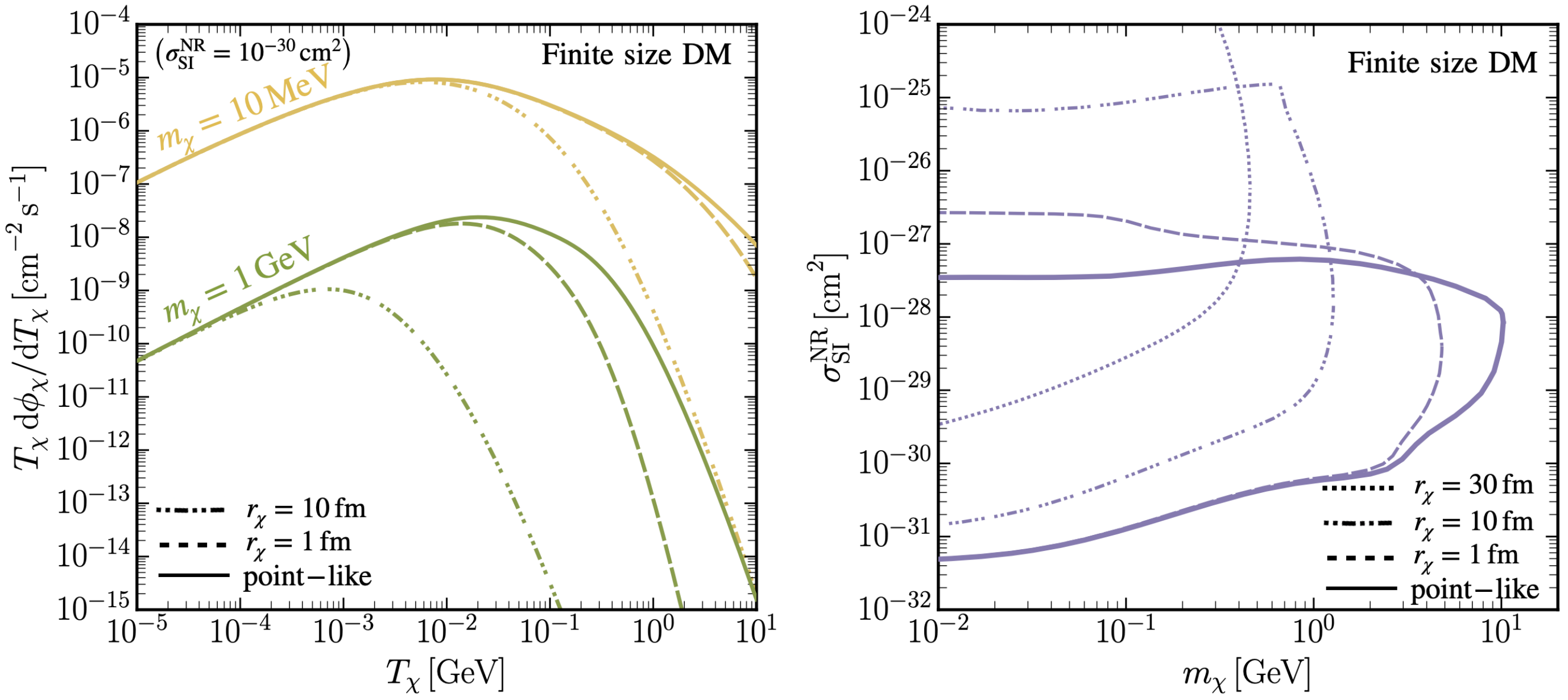}
\caption{{\it Left panel.} 
Solid lines show the CRDM flux before attenuation for a constant interaction
cross section, as in Fig.~\ref{fig:flux_ff}, for DM masses  $m_\chi=10$\,MeV and  $m_\chi=1$\,GeV. 
For comparison we indicate the corresponding CRDM flux for {\it finite size} DM,  
with  $r_\chi =1$\,fm (dotted) and $r_\chi=10$\,fm (dashed), for a cross section of  
$\sigma_{\rm SI}^{\rm NR}=10^{-30}\,\mathrm{cm}^2$. 
{\it Right panel.} Limits on the spin-independent DM-nucleon cross section, with line styles and colors matching 
those of the left panel. In particular, solid lines show the case of a constant scattering cross 
section and are identical to those displayed  in the left panel of Fig.~\ref{fig:constraints_constant}.
\label{fig:results_puffy}
}
\end{center}
\end{figure}

In a very similar fashion to what happens in the presence of a light mediator $\phi$, such a cross section features a sharp 
suppression for momentum transfers exceeding a `mass' scale $m_\phi\sim \sqrt{12}/r_\chi$. Sharper than in that case, in fact, as 
the suppression scales with 
a power of $Q^{-8}$ rather than just $Q^{-4}$. This is clearly visible in the left panel of Fig.~\ref{fig:results_puffy},
where we plot the expected CRDM flux for DM with a finite size, for various values of $m_\chi$ and $r_\chi$. 
For example, for $r_\chi=10$\,fm, we have 
$\sqrt{12}/r_\chi\sim68$\,MeV and the cutoff indeed appears at only slightly smaller values of $T_\chi$ than in the case
of the $100$\,MeV mediator displayed in Fig.~\ref{fig:results_scalar} (for $m_\chi=1$\,GeV). The slope above the cutoff, 
however, is twice as steep -- as expected from the $Q^{-8}$ suppression.

In the right panel of Fig.~\ref{fig:results_puffy} we show how the constraints on a constant DM-nucleon cross section
weaken when considering the situation where the DM particles themselves have a finite extent. Concretely,
for a DM radius of $r_\chi=1$\,fm ($r_\chi=10$\,fm) the maximal DM mass that can be probed decreases from
$\sim10$\,GeV to about $4.5$\,GeV ($1.1$\,GeV). The reduced CRDM flux for extended DM, cf.~the left panel
of the figure, also 
visibly weakens the lower bound on the exclusion region. At the same time, attenuation is also less efficient
for a given cross section in the non-relativistic limit (inelastic scattering still effectively cuts off the incoming CRDM
flux above $\sim$0.2\,GeV, explaining e.g.~the upper, almost horizontal boundary of the exclusion region in the
$r_\chi=10$\,fm case).
For $r_\chi\gtrsim1$\,fm, this starts to significantly enlarge the excluded region to higher cross sections.
On the other hand, it should be noted that for composite DM particles the interaction
cross section may not actually continue to drop as $Q^{-8}$ for very large momentum transfers,
as would be implied by Eq.~(\ref{eq:dipole}). At some point, instead, inelastic scattering events on the DM constituents 
will take over, in analogy to what we discussed for nuclei in section \ref{sec:inel}. This is particularly relevant 
if the DM constituents are themselves finite in size, in which case the upper boundaries of the
exclusion regions shown in Fig.~\ref{fig:results_puffy} would be overly optimistic for very large $r_\chi$.

Similar to the discussion in section \ref{sec:scalar}, a comparison of the limits shown in Fig.~\ref{fig:results_puffy}  
with complementary limits requires a re-scaling of $\sigma_{\rm SI}$ to a common reference cross section. Due to the
strong form factor suppression, this rescaling has an even larger effect than in the light mediator case; concretely,
instead of Eq.~(\ref{eq:rescale_scalar}), the rescaling of reported limits, $\sigma_{\rm SI}^p$, to those relevant 
for the Xenon-1T detector now takes the form
\be
 \tilde \sigma_{\rm Xe,SI}^p = \sigma_{\rm SI}^p\times 
 \left(\frac{Q^2_{\rm ref}+12/r_\chi^2}{Q^2_{\rm Xe,ref}+12/r_\chi^2}\right)^4\,.
\ee
Qualitatively, however, this does not change the lesson learned in the light mediator case: while limits from the CRDM component
can be weakened by increasing $r_\chi$, this will inevitably strengthen complementary bounds from cosmology. As a result,
we find once again an absolute upper bound on the cross section of about  
$\tilde \sigma_{\rm SI}\sim 3\cdot10^{-31}\,{\rm cm}^2$, independently of the DM mass
{\it and} size. Also in this case there is a small loophole to this statement if one is willing to assume that the thermalization
efficiency of CRESST is as small as 2\,\%: when tuning the size of the DM particles to $r_\chi\simeq10$\,fm,
we find that cross sections two orders of magnitude larger may in that case be viable for DM masses in a narrow range 
between around 1\,GeV and 2\,GeV.

\section{Hexaquarks: a viable baryonic dark matter candidate?}
\label{sec:hexaquark}

In section \ref{sec:m2} we discussed various generic situations where the amplitude for elastic scattering 
shows a significant dependence on the momentum transfer, and how this impacts the conclusions about 
whether a window of large scattering cross sections remains open or not. In this section we complement
those more model-independent considerations by taking a closer look at a specific DM candidate
in the GeV range, with relatively large nuclear interactions.
Concretely, it has been conjectured that a neutral (color-flavor-spin-singlet) bound state of six light quarks 
$uuddss$ may exist, and provide a plausible DM candidate that would evade all current
constraints despite its baryonic nature~\cite{Farrar:2002ic,Zaharijas:2004jv,Farrar:2017eqq, Farrar:2020zeo,Farrar:2022mih}. 
In particular, this \emph{sexaquark} $S$ (to be distinguished from a generic 6-quark state, often referred to as {\it hexaquark})
would form early enough to behave like standard cold DM during both big bang nucleosynthesis and recombination.
It would thus not be in conflict with the independent, and rather precise, measurements~\cite{Aver:2015iza,Planck:2018vyg} 
of the cosmological baryon density during these epochs.

Compared to the H-dibaryon that was suggested earlier~\cite{Jaffe:1976yi} and thoroughly studied both theoretically  and experimentally (see Refs.~\cite{Sakai:1999qm,Clement:2016vnl} for reviews), furthermore, the $S$ should be much 
more tightly bound, leading to weaker interactions with ordinary baryons and thus evading direct searches.  
Such a particle would be absolutely 
stable for  $m_S<m_D + m_e\simeq 1.88$\,GeV, and decay with a lifetime exceeding the age of the 
Universe for $m_S\lesssim 2\,$GeV~\cite{Farrar:2020zeo}. Determining its expected mass exactly, however,
is challenging; lattice simulations, for example, remain somewhat inconclusive (see, e.g., 
Refs.~\cite{NPLQCD:2010ocs,Inoue:2010es,NPLQCD:2012mex,Francis:2018qch} where the results for binding energies 
of the H-dibaryon state range from $\sim$17\,MeV to $\sim$75\,MeV relying, however, on unrealistically large quark masses).
Even if the sexaquark is stable on cosmological timescales, its relic abundance would generally be much smaller than
the observed DM abundance if one assumes that its interactions in the early universe are of the order of the 
strong force~\cite{Kolb:2018bxv,Gross:2018ivp}. If instead, one postulates much weaker interactions due to the assumed
compactness of the sexaquark, thermal equilibrium with the SM heat bath would
not be possible to maintain after the QCD phase transition and the correct DM abundance might be 
achieved -- in a region of parameter space claimed to evade all existing constraints~\cite{Farrar:2020zeo}.

Motivated by this intriguing possibility, for simplicity we will adopt the description of sexaquark interactions 
from Ref.~\cite{Farrar:2020zeo}, i.e.~we model the interaction with nucleons by the exchange 
of a vector meson. 
In particular, the relevant interaction terms with the flavour-neutral mixture of $\phi$ and $\omega$, denoted by $V$, are given by
\begin{equation}
\label{eq:laghexaquarks}
\mathcal{L}= V_\mu \left( i g_{S} S^\dagger {\mathop{\partial^\mu}^{\leftrightarrow}} S  + g_{p}\overline{p}\gamma^\mu p + g_{n}\overline{n}\gamma^\mu n\right),
\end{equation}
and we adopt the value $m_V = 1\,$GeV used in Ref.~\cite{Farrar:2020zeo} for our calculations.
The value of $g_n=g_p\sim 2.6\sqrt{4\pi}$ 
can be inferred from the literature on the one-boson-exchange model~\cite{Maessen:1989sx} although $\mathcal{O}(1)$ 
uncertainties can be expected here.\footnote{%
In particular, we note that modern analyses of low-energy baryon-baryon scattering consider processes beyond single 
meson exchange~\cite{Nagels:2015lfa}, and that baryon-baryon interactions can also be treated within the more 
systematic approach 
of chiral perturbation theory \cite{Weinberg:1990rz}. However, given the significant uncertainties on the sexaquark 
couplings we consider the one-boson-exchange approximation to be sufficient for our purposes.
}
The coupling $g_{S}$ is largely unknown, though simple scaling arguments suggest that
\begin{equation}
\alpha_{\mathrm{SN}} \equiv \frac{g_{S}g_{p}}{4\pi}
\end{equation}
is very roughly of the order of $\sim0.1$~\cite{Farrar:2020zeo}. Following that reference, we will treat $\alpha_{\mathrm{SN}}$
as a free parameter that we will generously vary in the interval $(10^{-3},10)$. 
Importantly however -- at least in this parameter range -- the DM relic abundance 
is independent of $\alpha_{\mathrm{SN}}$. Instead, the final abundance of $S$ is set by an independent coupling constant $\tilde{g}$~\cite{Farrar:2020zeo} 
that describes the (much weaker) sexaquark-breaking interactions within the effective description. This coupling does not directly enter the analysis presented here.

We treat the interaction of $V$ with nuclei similarly to that in section~\ref{sec:vector}, i.e.~we describe it by the effective 
Lagrangian~\eqref{eq:leff_vector} with the coherently enhanced, effective coupling $g_N$ given by Eq.~\eqref{eq:gN_coh}. 
For the elastic scattering cross section on nuclei we thus find
\begin{eqnarray}
\label{diffsig_full_hex}
\frac{d\sigma_{S N}}{d T_N}&=&\frac{\mathcal{C}^2 \sigma_{\rm SI}^\mathrm{NR}}{T_N^\mathrm{max}}
\frac{m_V^4}{(Q^2+m_V^2)^2}
\times G_N^2(Q^2) G_V^2(Q^2)\\
&&
\times\frac{1}{4s \mu_{S N}^2}
\left\{
\begin{array}{ll}
(s-\tfrac{1}{2}Q^2-m_N^2-m_\chi^2)^2& ~~\mathrm{for~scalar~}N\\
m_N^2Q^2-Q^2s+(s-m_N^2-m_S^2)^2&  ~~\mathrm{for~fermionic~}N
\end{array}
\right..\nonumber
\end{eqnarray}
Here, 
\begin{equation}
\sigma_{\rm SI}^{\rm NR} = \frac{16 \pi \alpha_{\mathrm{SN}}^2 \mu_{S p}^2}{m_V^4}
\end{equation}
is the scattering cross section on nucleons in the non-relativistic limit and $\mu_{S p}$ ($\mu_{S N}$) is the reduced mass of 
the sexaquark-nucleon (nucleus) system. 

Compared to the treatment in section~\ref{sec:vector}, we introduce an additional form factor $G_V$ related to the cutoff in the 
one-boson-exchange models. In this context, exponential cutoffs
\begin{equation}
G_V(Q^2) = e^{-\frac{Q^2}{\Lambda_V^2}}
\end{equation}
are mostly used and the cutoff mass $\Lambda_V$ is fitted to data (and can in principle differ for different meson exchange 
channels). For example, within the fit to data taking into account hyperon-nucleon interactions \cite{Maessen:1989sx}, these cutoff 
masses were found to range between 820\,MeV and 1270\,MeV. Since yet lower cutoff masses appear in related literature (e.g., 
down to 590\,MeV in \cite{Stoks:1996yj}), we generously vary $\Lambda_V$ between 500 and 1500\,MeV.
We note that for $\Lambda_V\gtrsim1500\,$MeV, CRDM limits become in fact independent of the cutoff scale.

\begin{figure}[t]
\begin{center}
\includegraphics[width=0.75\textwidth]{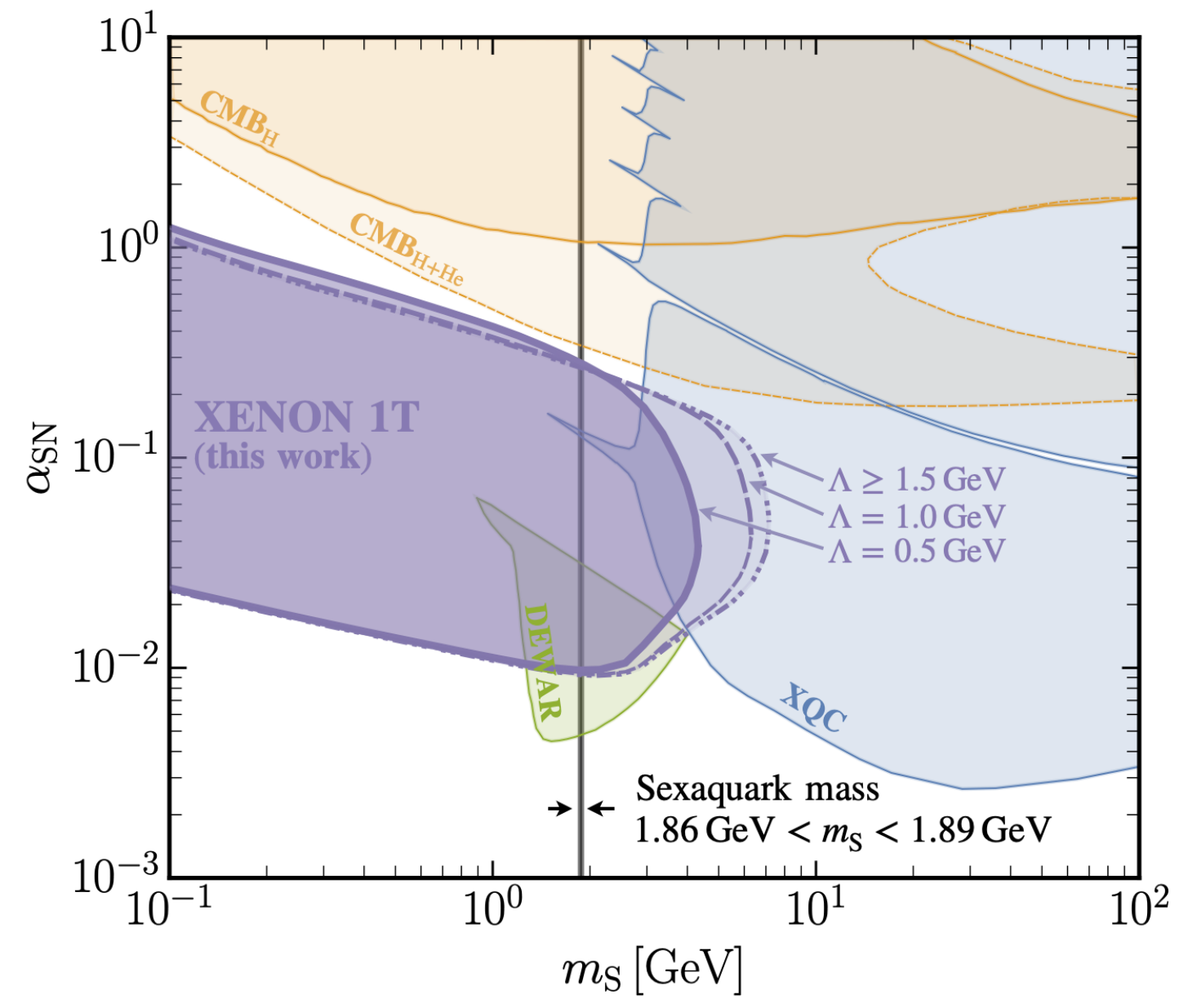}
\caption{Effective sexaquark coupling $\alpha_{\mathrm{SN}}$ vs.~sexaquark mass $m_S$.
The purple region shows the parameter range that is excluded by the analysis in this work, assuming that sexaquarks
make up all of the cosmologically observed DM;
different line styles correspond, as indicated, to cutoff masses $\Lambda_V/{\rm GeV}\in\{0.5, 1,1.5\}$
in the one-boson exchange approximation. All other constraints are, for easier comparison, directly
reproduced from Fig.~10 of Ref.~\cite{Farrar:2020zeo}, conservatively assuming an attractive Yukawa force between $S$
and nuclei. The thin vertical stripe corresponds to the mass range
where, according to that analysis, the sexaquark would be a viable DM candidate without being in conflict
with other particle physics observation, in particular the stability of deuterons based on SNO data~\cite{Bellerive:2016byv}. 
The upper end of that mass range may increase from 1890\,MeV to up to 2054\,MeV if sexaquark DM does not 
accumulate in the Earth at the level claimed in Ref.~\cite{Neufeld:2018slx}.
\label{fig:results_sexaquark}
}
\end{center}
\end{figure}

In Fig.~\ref{fig:results_sexaquark} we show the parameter space in the $\alpha_{\mathrm{SN}}$ vs.~$m_S$ plane where sexaquark DM
is excluded because of the irreducible CRDM component. For a better direct comparison, we also indicate the preferred
mass range according to Ref.~\cite{Farrar:2020zeo}, along with the complementary limits presented in that analysis. 
From this figure, it is clear that our new limits close a significant part of the viable parameter region where sexaquarks could be
the dominant DM component -- even without taking into account the CRESST results. 
In particular, we note that the Lyman-$\alpha$ limits~\cite{Rogers:2021byl}  
shown in figures~\ref{fig:constraints_constant} and \ref{fig:limits_scalar} were presented subsequent to the analysis of
Ref.~\cite{Farrar:2020zeo} and are significantly stronger than the CMB limits indicated in Fig.~\ref{fig:results_sexaquark}.
The apparently open window at $\alpha_{\mathrm{SN}}\sim0.3$ is thus also robustly excluded.
On the other hand, a small open window remains for $\alpha_{\mathrm{SN}}\lesssim 4\cdot10^{-3}$. While not being
in conflict with the DM abundance, as explained above, we recall that 
such values of  $\alpha_{\mathrm{SN}}$ are somewhat smaller than intrinsically expected.

Let us, finally, briefly comment on the fact that the DM-nucleon scattering cross section can, strictly speaking,
only be calculated perturbatively in the Born limit, $\alpha_{\mathrm{SN}} \mu_{\chi N}\lesssim m_V$. Outside this regime,
non-relativistic scattering in a Yukawa potential exhibits parametric resonances where the scattering amplitude
is significantly enhanced or suppressed. This non-perturbative 
effect is well-known from the self-scattering of cold DM in the presence
of light mediators~\cite{Tulin:2013teo}, and it is the origin of the resonant structure in the complementary limits from 
Ref.~\cite{Farrar:2020zeo} that is visible in Fig.~\ref{fig:results_sexaquark}. For our CRDM limits, on the other hand, 
this additional complication does not arise because such non-perturbative corrections are 
largely irrelevant for relativistic scattering; in fact, 
already for the typical velocities during the freeze-out process of thermally produced DM, $v_\chi\sim0.3$, the impact
is strongly suppressed~\cite{Tulin:2013teo}. The CRDM limits are thus also robust w.r.t.~underlying model assumptions 
such as whether the force mediated by the Yukawa potential is attractive or repulsive.

\section{Summary and Conclusions}
\label{sec:conclusions}

For sizeable elastic scattering rates between DM and nuclei there is an irreducible relativistic
component of the flux of DM particles arriving at Earth. This extends the sensitivity of conventional
direct detection experiments both to sub-GeV masses and to scattering cross sections above the limit
set by a too efficient attenuation of the DM flux on the way to the detector. While such large scattering cross 
sections are also constrained by complementary probes from astrophysics and cosmology, it has repeatedly 
been pointed out that there might be an open window of relatively strongly interacting DM with a mass in the
ballpark of $\sim1$\,GeV.

We find that the CRDM component in the DM flux generically closes this window, under rather minimal 
assumptions. In order to arrive at this conclusion, we included in our analysis a detailed treatment
of the inelastic scattering of DM off nuclei (section \ref{sec:inel}). We demonstrate that this provides 
an important additional stopping channel for CRDM particles on their way to direct detection facilities
-- unlike for non-relativistic DM, where only elastic scattering is relevant.
We also investigated the extent to which a possible energy or momentum-transfer dependence of the cross section
could weaken our general conclusions. 
For this purpose, we considered {\it i)} a class of simplified models where the scattering with nuclei is mediated by a light
scalar (section \ref{sec:scalar}) or vector (section \ref{sec:vector})
particle, as well as {\it ii)} situations where DM particles cannot be described as being point-like (section \ref{sec:puffy}). 
In all these cases, the additional momentum-transfer dependence indeed weakens the limits from
direct detection -- which however is compensated for by a corresponding strengthening of complementary
limits, in particular from cosmology. In combination, these limits stringently constrain the 
possibility of cross sections larger than a few times $10^{-31}\,{\rm cm}^2$, over a wide range of DM masses. Interestingly, this is
largely independent of underlying modelling assumptions such as the mass of new mediator particles or
the DM particles' radius.

Finally, an exotic QCD bound state that is produced well before BBN, has repeatedly been put forward as 
a potential DM candidate. While it is theoretically unclear whether such states could actually exist, 
adding to significant experimental constraints, it is certainly an intriguing
idea to have a `baryonic' DM candidate that would in fact evade the strong
evidence from BBN and CMB against this possibility. However, cosmic-ray upscattering
of such particles leads to stringent new constraints that have not previously been pointed out
in this context. For the concrete case of stable sexaquark DM, as discussed in 
section~\ref{sec:hexaquark}, we find that the parameter space giving the correct cosmological
abundance is strongly pressured.

For the analysis performed in this work we used the numerical tool \ds~\cite{Bringmann:2018lay} 
to compute CRDM fluxes and limits. In doing so we significantly expanded the general numerical routines
provided therein, adding in particular inelastic scattering, the contribution 
from CRs beyond $p$ and He, and an updated treatment of nuclear form factors in the context of
CRDM attenuation.   These updates
will be included in the next public release 
of the code.

\acknowledgments
We thank Timon Emken and Florian Reindl for enlightening discussions about the thermalization efficiency of the
CRESST experiment, and Felix Kahlhoefer for insightful comments on how to map nucleon to nuclear cross sections. We further 
thank Assumpta Pare\~{n}o, Gilberto Colangelo and Urs Wiedemann for comments related to the hexaquark state. TB warmly 
thanks the Albert Einstein Institute in Bern, and the CERN Theoretical Physics Department, 
for support and hospitality during the preparation of this manuscript. JA is supported through the research program ``The Hidden 
Universe of Weakly Interacting Particles" with project number 680.92.18.03 (NWO Vrije Programma), which is partly financed by 
the Nederlandse Organisatie voor Wetenschappelijk Onderzoek (Dutch Research Council). HK was supported by the ToppForsk-UiS Grant No. PR-10614 and by the Swiss National Science Foundation (SNSF) under grant 200020B-188712.

\begin{table}[]
	\resizebox{\textwidth}{!}{
	\begin{tabular}{@{}llllll@{}}
	\toprule
	\multicolumn{2}{l}{\hspace{-0.2cm}\textbf{\&neutrino\_induced}} & \multicolumn{2}{l}{\textbf{\&input}}               & \multicolumn{2}{l}{\textbf{\&nl\_dSigmadElepton}} \\
	\texttt{process\_ID}               & 3           & \texttt{eventtype}                     & 5                & \texttt{enu}                  & $T_\chi$             \\
	\texttt{flavor\_ID}                & 2           & \texttt{numEnsembles}                  & 100              & \texttt{elepton}              & $0.005 T_\chi$       \\
	\texttt{nuXsectionMode}            & 2           & \texttt{numTimeSteps}                  & 0                & \texttt{delta\_elepton}       & $\Delta E_\ell$      \\
	\texttt{nuExp}                     & 0           & \texttt{num\_Energies}                 & 50               & \multicolumn{2}{l}{\textbf{\&target}}                \\
	\texttt{includeQE}                 & T/F         & \texttt{num\_runs\_sameEnergy}         & 1                & \texttt{Target\_A}            & $A$                  \\
	\texttt{includeDELTA}              & T/F         & \texttt{delta\_T}                      & 0.2              & \texttt{Target\_Z}            & $Z$                  \\
	\texttt{includeRES}                & T/F         & \texttt{localEnsemble}                 & T                & \multicolumn{2}{l}{\textbf{\&initDensity}}           \\
	\texttt{path\_To\_Input}           & \texttt{/path/to/buuinput} & \texttt{include1pi}     & F                & \texttt{densitySwitch}        & 2                    \\
	\texttt{includeDIS}                & T/F         & \multicolumn{2}{l}{\textbf{\&neutrinoAnalysis}}           & \multicolumn{2}{l}{\textbf{\&initPauli}}             \\
	\texttt{2p2hQE}                    & F           & \texttt{XSection\_analysis}            & T                & \texttt{pauliSwitch}          & 2                    \\
	\texttt{include2p2hDelta}          & F           & \texttt{detailed\_diff\_output}        & F                &  													\\
    \texttt{include2pi}                & F           &                                        &                  &  													\\ \bottomrule
	\end{tabular}
	}
	\caption{Settings choices for running \texttt{GiBUU} to study neutral current neutrino scattering.
	We also enforced a logarithmic binning in the outgoing lepton energy, by changing the variable assignment 
	of \texttt{dElepton} from $E_\ell \rightarrow E_\ell + \Delta E_\ell$ to $E_\ell \rightarrow (1 + \Delta E_\ell) E_\ell$.
\label{tab:gibuu}
	}
	\end{table}

\bibliographystyle{JHEP_improved}
\bibliography{biblio.bib}

\end{document}